\documentclass[conference]{IEEEtran}
\IEEEoverridecommandlockouts

\usepackage[pdftex]{graphicx}

\usepackage{amsmath,amsfonts,amssymb,amsthm}
\usepackage{bbm}

\newtheorem{definition}{Definition}

\DeclareMathOperator*{\argmin}{argmin}

\usepackage{algorithm}
\usepackage{algorithmicx}
\usepackage[noend]{algpseudocode}

\usepackage{url}
\usepackage{hyperref}

\usepackage{booktabs}
\usepackage{makecell}
\usepackage{multirow}

\usepackage{adjustbox}
\usepackage{xcolor}
\usepackage[many]{tcolorbox}
\usepackage{soul}
\usepackage{subfigure}
\usepackage{pifont}
\usepackage{mdframed}

\newcommand{\para}[1]{\vspace{0pt}{\noindent\textbf{#1}}}

\usepackage{tikz}

\def\BibTeX{{\rm B\kern-.05em{\sc i\kern-.025em b}\kern-.08em
    T\kern-.1667em\lower.7ex\hbox{E}\kern-.125emX}}

\begin{document}

\title{Reconstruction of Differentially Private Text Sanitization via Large Language Models

\thanks{\textsuperscript{$\spadesuit$}Shuchao Pang and Zhigang Lu contributed equally. Yongbin Zhou is the corresponding author.}
\thanks{This work is supported by the National Natural Science Foundation of China (Grant No.62206128, No.U2336205), National Key Research and Development Program of China under (Grant No.2023YFB2703900). This work was partially done, when Zhigang Lu was a lecturer with James Cook University. Zhigang Lu is financially supported by Western Sydney University Start-up Funds.}
}

\author{\IEEEauthorblockN{Shuchao Pang\IEEEauthorrefmark{1}\textsuperscript{$\spadesuit$}, 
Zhigang Lu\IEEEauthorrefmark{2}\textsuperscript{$\spadesuit$}, 
Haichen Wang\IEEEauthorrefmark{1},
Peng Fu\IEEEauthorrefmark{3},
Yongbin Zhou\IEEEauthorrefmark{1},
Minhui Xue\IEEEauthorrefmark{4}\IEEEauthorrefmark{5}} 
\IEEEauthorblockA{\IEEEauthorrefmark{1}Nanjing University of Science and Technology, China
}
\IEEEauthorblockA{\IEEEauthorrefmark{2}Western Sydney University, Australia
}
\IEEEauthorblockA{\IEEEauthorrefmark{3}Institute of Information Engineering, Chinese Academy of Sciences, China
}
\IEEEauthorblockA{\IEEEauthorrefmark{4}CSIRO's Data61, Australia
}
\IEEEauthorblockA{\IEEEauthorrefmark{5}Responsible AI Research (RAIR) Centre, The University of Adelaide, Australia}
}

\maketitle

\begin{abstract}
Differential privacy (DP) is the de facto privacy standard against privacy leakage attacks, including many recently discovered ones against large language models (LLMs). However, we discovered that LLMs could reconstruct the altered/removed privacy from given DP-sanitized prompts. We propose two attacks (black-box and white-box) based on the accessibility to LLMs and show that LLMs could connect the pair of DP-sanitized text and the corresponding private training data of LLMs by giving sample text pairs as instructions (in the black-box attacks) or fine-tuning data (in the white-box attacks). To illustrate our findings, we conduct comprehensive experiments on modern LLMs (e.g., LLaMA-2, LLaMA-3, ChatGPT-3.5, ChatGPT-4, ChatGPT-4o, Claude-3, Claude-3.5, OPT, GPT-Neo, GPT-J, Gemma-2, and Pythia) using commonly used datasets (such as WikiMIA, Pile-CC, and Pile-Wiki) against both word-level and sentence-level DP. The experimental results show promising recovery rates, e.g., the black-box attacks against the word-level DP over WikiMIA dataset gave $72.18\%$ on LLaMA-2 (70B), $82.39\%$ on LLaMA-3 (70B), $75.35\%$ on Gemma-2, $91.2\%$ on ChatGPT-4o, and $94.01\%$ on Claude-3.5 (Sonnet). More urgently, this study indicates that these well-known LLMs have emerged as a new security risk for existing DP text sanitization approaches in the current environment.
\end{abstract}


%
\section{Introduction}
\label{sec:introduction}

State-of-the-art large language models (LLMs), such as ChatGPT~\cite{achiam2023gpt} and LLaMA~\cite{touvron2023llama}, have been applied in numerous real-world applications due to their unprecedented capabilities, achieved through training billions of parameters on vast amounts of text from the Internet~\cite{gao2020pile}. However, several studies~\cite{carlini2021extracting,huang2022large,nasr2023scalable,sha2024prompt,hui2024pleak,agarwal2024investigating,lukas2023analyzing,mattern2023membership,shi2024detecting,mireshghallah2022quantifying,zhang2024effective} have reported privacy leaks due to interactive communication between external users or adversaries and LLMs. For instance, \cite{huang2022large,nasr2023scalable,lukas2023analyzing} exploit LLMs to extract sensitive information from their training data. Others like \cite{mattern2023membership,shi2024detecting,mireshghallah2022quantifying} demonstrate membership inference attacks (MIAs) against LLMs' training datasets. More recently, \cite{zhang2024effective,sha2024prompt,agarwal2024investigating} show that adversaries can reconstruct original prompts based solely on LLMs' responses.

Unlike existing studies that focus on the ``current'' privacy issues of LLMs, this paper studies ``future'' privacy concerns. Our motivation stems from a straightforward yet critical observation: in reality, the boundaries of privacy are not static. Information that was previously in the public domain and used in the training data of LLMs may later become private. For instance, Article 17 of the General Data Protection Regulation (GDPR) specifies the practical circumstance for the \emph{right to be forgotten}, where ``The personal data is \emph{no longer necessary} for the purpose an organization originally collected or processed it''~\cite{gdpr}. The future needs for privacy preservation require LLMs to provide answers based on up-to-date knowledge, aligning with the latest privacy standards.

\begin{figure}[!t]
    \centering
    \includegraphics[width=\linewidth,page=1]{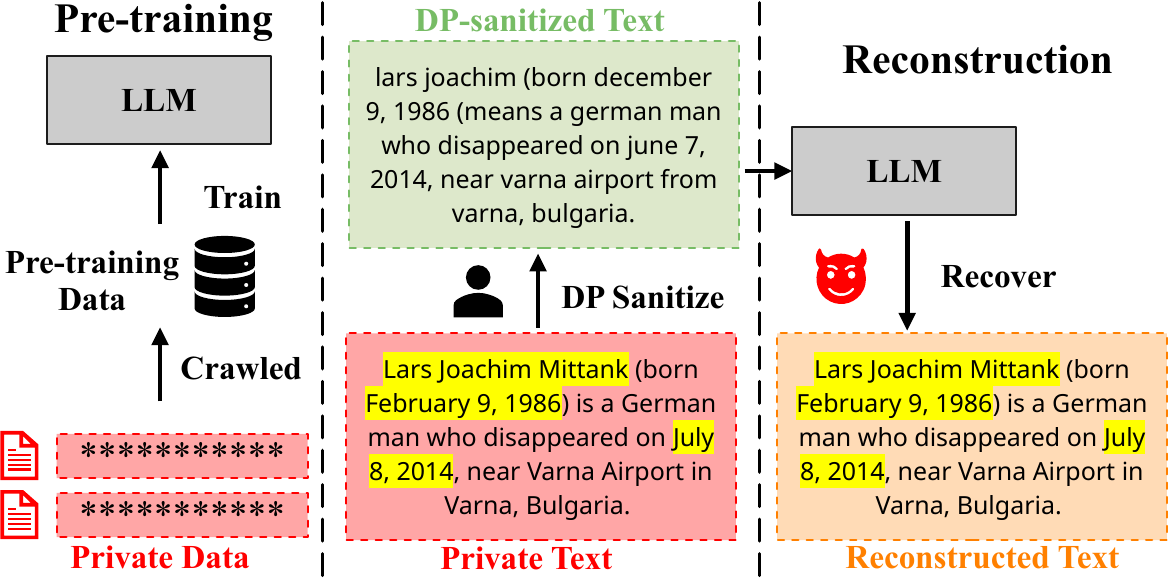}
    \caption{Overview of reconstruction privacy from given DP-sanitized prompts using LLMs.}
    \label{fig:overview}
\end{figure}

A seemingly feasible solution to meet the future privacy needs, without re-training pre-existing LLMs, involves providing differentially private~\cite{dwork2006calibrating} information as the prompts through an interactive retrieval-augmented generation (RAG) process. This allows the LLMs to provide customized responses based on the privacy-preserved information without accessing private details, such as dates, names, numbers, and addresses. More importantly, the randomness of DP, making it a one-way function, further ensures that it is (nearly) impossible to deduce the protected privacy information from the sanitized text (potentially leaked by the prompts leakage attacks). To pre-process/sanitize the original (private) text, we can either implement the word level DP~\cite{feyisetan2019leveraging,feyisetan2020privacy} or the sentence/document level DP~\cite{krishna2021adept,igamberdiev2023dp,mattern2022limits,utpala2023locally}.

However, in this work, our attacks against DP-sanitization discovered that an adversary can take advantage of the updated LLMs (meeting future privacy needs) to invert non-invertible DP outputs. Specifically, if the private text used during the training of LLMs is then sanitized using DP and fed back to the LLMs as prompts, LLMs can reconstruct the original private information.

To illustrate such a capability, we consider both black-box and white-box adversaries, where the black-box access allows the adversaries to prompt the LLM with a prefix and obtain the probability distribution of the next token via API, whereas the white-box access lets the adversaries learn and modify the model parameters of a given LLM. Based on how we interact with the LLMs, i.e., instruction-based tuning and fine-tuning, we designed two attacks accordingly - black-box instruction-based attacks and white-box fine-tuning-based attacks. In particular, in the black-box instruction-based attacks, we design queries to make instruction-tuned LLMs respond with their memorized content based on the sanitized text for the adversary. In the white-boxed fine-tuning-based attack, the adversary adapts pre-trained LLMs to the reconstruction by fine-tuning it on an auxiliary dataset for a DP text sanitization approach under a specific privacy budget. Note that our attacks can be adapted to any LLM. Fig.~\ref{fig:overview} depicts an overview of the reconstruction of the private information given DP-sanitized prompts.

We concluded the aforementioned finding against differential privacy by conducting comprehensive experiments on modern LLMs (LLaMA-2, LLaMA-3, ChatGPT-3.5, ChatGPT-4, ChatGPT-4o, Claude-3, Claude-3.5, OPT, GPT-Neo, GPT-J, Gemma-2, and Pythia) using commonly used datasets (WikiMIA, Pile-CC, and Pile-Wiki) against both word-level and sentence-level DP. Our experimental results demonstrated varying levels of success based on model parameters, privacy budgets, and dataset characteristics. For example, let the privacy budget be $12$ (which is a commonly used privacy budget in real life to keep the semantics at a reasonable level~\cite{census}), the success of reconstruction through the black-box instruction-based attacks against the word-level DP over WikiMIA dataset are $72.18\%$ on LLaMA-2 (70B), $82.39\%$ on LLaMA-3 (70B), $75.35\%$ on Gemma-2, $91.2\%$ on ChatGPT-4o, and $94.01\%$ on Claude-3.5 (Sonnet). Since LLMs do not have long-term memory~\cite{vaswani2017attention,brown2020language}, our findings further argue that the LLMs' ability to reconstruct private information is due to their exposure to the original (unsanitized) data during training, which might have enhanced their precision in recovering certain data. 

Our contributions can be summarized as follows.

\begin{itemize}
    \item We first discover LLMs are capable of reconstructing privacy from differentially private outputs.
    \item We propose two novel attacks based on the accessibility to large language models - black-box instruction-based attacks against LLMs giving no model parameter access and white-box fine-tuning-based attacks against LLMs giving model parameter access.
    \item We provide a novel metric to evaluate the reconstruction attacks on both word level and sentence level sanitization.
    \item We conduct comprehensive experiments on well-known LLMs and commonly used dataset to show the effects of our attacks against both word-level DP and sentence-level DP.
\end{itemize}

%


\section{Background}\label{sec:background}

This section briefly introduces the existing DP text sanitization approaches and scope of privacy. We give the summary of notations used in this paper in Tab.~\ref{tab:summary_of_notation} and the introduction to language models in Appx.~\ref{app:lm}.
\begin{table}[th]
    \centering
    \caption{Summary of Notations.}
    \label{tab:summary_of_notation}
    \begin{adjustbox}{max width=\linewidth}
         \begin{tabular}{p{0.42\linewidth}|p{0.55\linewidth}}
            \toprule
            \textbf{Notation} & \textbf{Definition} \\
            \midrule
            $\mathcal{T}$ & Training algorithm \\
            $\mathcal{V}$ & Vocabulary space \\   
            $\mathcal{X}$ & Sequence distribution \\
            $\theta$ & Model parameters \\
            ${X}\sim\mathcal{X}$ & Sample a pre-training dataset from $\mathcal{X}$ \\
            ${X}_{\text{aux}}$ & Auxiliary dataset for fine-tuning \\
            $\mathcal{P}$ & DP text sanitization approach \\
            $\mathcal{P}^{-1}$ & Reconstruction procedure for DP sanitization mechanism \\
            $\boldsymbol{x}=\{x_1,\dots,x_n\}\in\mathcal{X}$ & Sequence with $n$ tokens in the sequence distribution \\
            $\boldsymbol{z}\leftarrow\textsc{Concatenate}(\boldsymbol{x},\boldsymbol{y})$ & Concatenate sequence $\boldsymbol{x}$ and $\boldsymbol{y}$ \\
            $\boldsymbol{y}\leftarrow\textsc{GenPrompt}(\boldsymbol{p}_{T},\boldsymbol{x})$ & Construct prompt using the template $\boldsymbol{p}_{T}$ and sequence $\boldsymbol{x}$ \\
            $\boldsymbol{x}'\leftarrow\textsc{Generate}(\boldsymbol{p},\theta)$ & Text generation using the LM with parameter $\theta$ for prefix $\boldsymbol{p}$ \\ 
            $\tilde{\boldsymbol{x}}\leftarrow\textsc{Sanitize}(\boldsymbol{x},\mathcal{P},\epsilon)$ & Text sanitization from private text $\boldsymbol{x}$ to sanitized text $\tilde{\boldsymbol{x}}$ using DP text sanitization mechanism $\mathcal{P}$ under privacy budget $\epsilon$ \\
            $\hat{\boldsymbol{x}}\leftarrow\textsc{Reconstruct}(\tilde{\boldsymbol{x}},\mathcal{P}^{-1})$ & Text sanitization inversion from sanitized text $\tilde{\boldsymbol{x}}$ to reconstructed text $\hat{\boldsymbol{x}}$ using reconstruction attack $\mathcal{P}^{-1}$ \\
            \bottomrule
        \end{tabular}
    \end{adjustbox}
\end{table}

\subsection{Differentially Private Text Sanitization}
Differential Privacy~\cite{dwork2006calibrating} (DP) provides privacy guarantees for every single record in a dataset by adding extra noise into query results to limit the influence of each record. In practice, DP has become the de facto standard of privacy definition for machine learning algorithms. Deferentially private text sanitization is commonly applied in text anonymization tasks. There are two types of well-established DP text sanitization approaches: word-level DP~\cite{feyisetan2019leveraging,feyisetan2020privacy} and document/sentence-level DP~\cite{krishna2021adept,igamberdiev2023dp,mattern2022limits,utpala2023locally}. We give the detailed description of the word-level DP and the sentence-level DP in Appx.~\ref{app:dp}. Alg.~\ref{alg:word_level_mechanism} and Alg.~\ref{alg:paraphrasing_based_mechanism} present the outline of the world- and sentence-level methods, respectively. Tab.~\ref{tab:existing_text_sanitization_methods} lists the summary of differentially private text sanitization methods. 

\begin{algorithm}[th]
    \caption{Word-level DP~\cite{feyisetan2020privacy}.}
    \begin{algorithmic}[1]
        \Require private text $\boldsymbol{x}=\{x_1,x_2,\dots,x_n\}$, privacy budget $\epsilon$, pre-trained word embedding model $\phi$
        \Ensure sanitized text $\tilde{\boldsymbol{x}}=\{\tilde{x}_{1},\tilde{x}_{2},\dots,\tilde{x}_{n}\}$
        \For{${i}\leftarrow{1},{2},\dots,{n}$}
            \State Obtain embedding $\phi_i\leftarrow\phi(x_i)$
            \State Sample noise $\boldsymbol{z}\sim{p}_{\epsilon}(\boldsymbol{z})$
            \State Perturb embedding $\tilde{\phi}_{i}\leftarrow\phi_{i}+\boldsymbol{z}$
            \State Obtain sanitized word $\tilde{x}_{i}\leftarrow\argmin_{v\in\mathcal{V}}\|\phi(v)-\tilde{\phi}_{i}\|$
        \EndFor
    \end{algorithmic}
    \label{alg:word_level_mechanism}
\end{algorithm}

\begin{algorithm}[th!]
    \caption{Sentence-level DP (based on paraphrasing)~\cite{mattern2022limits,igamberdiev2023dp}.}
    \begin{algorithmic}[1]
        \Require private text $\boldsymbol{x}$, temperature $T$, LM with parameters $\theta$, output length $m$, prompt template $\boldsymbol{p}_{T}$, clipping constant $C$, vocabulary $\mathcal{V}$
        \Ensure sanitized text $\tilde{\boldsymbol{x}}$
        \State Construct prompt $\boldsymbol{y}\leftarrow\textsc{GenPrompt}(\boldsymbol{p}_{T},\boldsymbol{x})$
        \State $\boldsymbol{x}'\leftarrow\emptyset$
        \For{${i}\leftarrow{1},{2}\dots,{m}$}
            \State Compute logit $\boldsymbol{u}$ with LM using prompt $\boldsymbol{y}$
            \State Clipping logit $\bar{\boldsymbol{u}}\leftarrow\textsc{Clip}(\boldsymbol{u},C)$
            \State Obtain distribution $\mathbb{P}\leftarrow\textsc{Softmax}(\boldsymbol{\bar{u}}, T)$
            \State Sample the next token ${v}$ from $\mathcal{V}$ using $\mathbb{P}$
            \State $\tilde{\boldsymbol{x}}\leftarrow\tilde{\boldsymbol{x}}\cup\{v\}$, $\boldsymbol{y}\leftarrow\boldsymbol{y}\cup\{v\}$
        \EndFor
    \end{algorithmic}
    \label{alg:paraphrasing_based_mechanism}
\end{algorithm}

For texts with different lengths, the word-level DP can not provide privacy guarantees for two texts of the same length because it outputs the fixed-length text. Nevertheless, the sentence-level DP can adapt to texts of varying lengths, with a linear increase in privacy budget as the length of the output text grows. For the privacy definition, based on metric DP, the distance used in the word-level DP mainly reflects word differences within the text but ignores semantic differences. In contrast, the sentence-level DP follows the Local DP (LDP), where any two texts are adjacent, making it hard to achieve a practical privacy budget. For the implementation of DP, the word-level DP introduces perturbation at the word level, and the sentence-level DP perturbs the semantic information of the text. Note that with a small privacy budget, the word-level DP adds a large-scale noise to the embeddings of the original words, significantly changing the word types and resulting in low utility. Tab.~\ref{tab:example_of_word_levl_and_sentence_level} shows an example of the word-level DP (Madlib under $\epsilon=8$) and the sentence-level DP (DP-Prompt under $T=2.0$) for text sanitization.


\begin{table}[th]
    \centering
    \caption{Summary of differentially private text sanitization methods.}
    \begin{adjustbox}{max width=\linewidth}
        \begin{tabular}{cccc}
            \toprule
            Type & Method & Privacy Definition & Privacy Mechanism \\
            \midrule
            \multirow{2}{*}{Word-level} & MadLib~\cite{feyisetan2020privacy} & Metric DP & Laplace \\
                                        & TEM~\cite{carvalho2023tem} & Metric DP & Truncated Exponential \\
            \midrule
            \multirow{4}{*}{Sentence-level} & ADePT~\cite{krishna2021adept} & LDP & Laplace/Gaussian \\
                                            & DP-BART~\cite{igamberdiev2023dp} & LDP & Laplace/Gaussian \\
                                            & Paraphraser~\cite{mattern2022limits} & LDP & Exponential \\
                                            & DP-Prompt~\cite{utpala2023locally} & LDP & Exponential \\
            \bottomrule
        \end{tabular}
    \end{adjustbox}
    \label{tab:existing_text_sanitization_methods}
\end{table}

\begin{table*}[th]
    \centering
    \caption{Example of the word-level DP and the sentence-level DP.}
    \begin{adjustbox}{max width=\linewidth}
        \begin{tabular}{l|p{0.8\linewidth}}
            \toprule
            \textbf{Type} & \makecell[c]{\textbf{Example}} \\
            \midrule
            \textbf{Original} & {Well, I'm Tony. I live in Taichung, Taiwan, but I come from Nova Scotia, Canada. Like most expatriates here, I teach English to pay the bills. I'm married to a Taiwanese woman and plan on staying here. I started gaming around the time the Wilderness Survival Guide was released.} \\
            \midrule
            \textbf{Word-level} & {making, i'm tony has i live in taichung, taiwan, but know want from nova scotia met canada. like ever expatriates here, i teach english it pay in cards and my'm married then a taiwanese babies it plan on anxious informed end i started gaming watch sense times vast wilderness aiming guide although released alone}\\
            \midrule
            \textbf{Sentence-level} & {Taiwan, but I originally hail from Nova Scotia, Canada. Similar to many foreigners here, I earn income by teaching the English language. I am happily married to a Taiwanese national and intend to establish permanent residence in Taichung. I began my gaming hobby during the time when the Wilderness Survival Guide was newly published.} \\
            \bottomrule
        \end{tabular}  
    \end{adjustbox}
    \label{tab:example_of_word_levl_and_sentence_level}
\end{table*}

\subsection{Personally Identifiable Information}

Personally Identifiable Information (PII) refers to any data that can be used to identify a specific individual. PII is crucial for privacy and security, as its misuse can lead to identity theft, financial loss, and privacy violations. Pil{\'a}n et al.~\cite{pilan2022text} category PII to direct identifier when one data can re-identify an individual and quasi-identifier when identifiers combined with others can re-identify an individual. This includes direct identifiers like names, phone numbers, and IP addresses, and quasi-identifiers like birth date, gender, and postal code. Named Entity Recognition (NER) is a technology that locates and classifies named entities in text into predefined categories (e.g., names, locations, times). NER is crucial in various natural language processing applications, including search engines, question-answering systems, and content categorization. State-of-the-art NER (e.g., Flair~\cite{akbik2019flair}, NLTK~\cite{bird2009natural}, and SpaCY~\cite{honnibal2020spacy}) classify sequences using a Transformer neural networks. In our case, we leverage NER to tag PII for the given text because it is challenging for a large dataset. In particular, Tab.~\ref{tab:pii_classes_meaning_and_example} provides the semantics and examples of classes for grouping using NER in our work. Finally, it is noteworthy that in our experiments we did not store any PII as auxiliary knowledge for our attacks.
\begin{table}[th]
    \centering
    \caption{Semantics and Examples of PII Classes.}
    \label{tab:pii_classes_meaning_and_example}
    \begin{tabular}{l|l|p{0.35\linewidth}}
        \toprule
         \textbf{Name} & \textbf{Semantics} & \textbf{Examples} \\
         \midrule
         cardinal & cardinal value & 181 \\
         date & date value & 2014 \\
         event & event name & Hurricane Ana \\
         fac & building name & the Stade de France \\
         gpe & geo-political entity & Egypt \\
         language & language name & Spanish \\
         law & law name & The Local Audit and Accountability Act 2014 \\
         loc & location name & the Central Pacific \\
         money & money name & €375 \\
         norp & affiliation & Conservatives \\
         ordinal & ordinal value & second \\
         org & organization name & Muslim Brotherhood \\
         percent & percent value & 42.1\% \\
         person & person name & Mohamed Badie \\
         product & product name & Ziyuan I-04 \\
         quantity & quantity name & 82 metres \\
         time & time value & 2:02:57 \\
         work of art & name of work of art & The BRW Rich 200, 2014 \\
         \bottomrule
    \end{tabular}
\end{table}

%
\section{Problem Formulation}\label{sec:problem_setting}

This section introduces the research question and the threat model, including the attacker's capabilities and the attack target.

\subsection{Problem Statement}

Recall our motivating example of the ``future'' privacy issues in Sec.~\ref{sec:introduction}. Since the scope of privacy might not be fixed, portions of the training data used for existing LLMs could potentially become private in the future, even if they were not initially considered as such.

Among all privacy enhance techniques, differential privacy (DP)~\cite{dwork2006calibrating} is an ideal solution to address such ``future'' privacy issues by using DP-sanitized text as prompts to update the knowledge base of a trained LLM. This allows the updated LLM to use the privacy-preserved knowledge instead of the original private information when answering external requests. Thanks to the rigorous privacy guarantee provided by DP, the existing prompts leakage attacks~\cite{zhang2024effective,sha2024prompt,agarwal2024investigating} can only reveal the DP-sanitized prompts but not the original ones.

Nevertheless, take a reasonable assumption where the DP-sanitized prompts should have similar semantics to the original text, a question remains open: \emph{In our future privacy circumstance, given DP-sanitized prompts, could LLMs recover the original privacy of the sanitized prompts?}

\subsection{Threat Model}

\para{Adversary's Capabilities.} We consider that adversaries either have black-box or white-box access to trained LLMs. For black-box access, an adversary can prompt the LLM with a prefix and obtain the probability distribution of the next token via paid APIs, e.g., ChatGPT and Claude; whereas the white-box adversaries have an auxiliary dataset following the same distribution as the private training dataset and the access to modify the model parameter $\theta$ of the target LLMs, such as OPT, Pythia, GPT-Neo and GPT-J. We further assume that the same as other DP approaches, the DP-sanitized prompts, the privacy budget and the DP approach (word-level DP or sentence-level DP) are publicly available~\cite{apple2022diff}. Note that the difference between the DP-sanitized text and the original text is determined by the privacy budget and the DP approach.

\para{Adversary's Target.} The adversaries aim to generate a reconstructed version of the original text corresponding to the DP-sanitized text. Prior work~\cite{carlini2023quantifying} shows that only $1\%$ of the pre-training dataset is memorized by some LLMs, indicating that generating verbatim memorization for LLMs directly is challenging. Therefore, compared with the previous works~\cite{nasr2023scalable,carlini2023quantifying} that focus on memorization of the pre-training dataset in LLMs, our goal is to investigate how to generate sensitive information rather than the verbatim sequence in the training data. 
\begin{definition}[Reconstruction]
    Let a language model having model parameters $\theta$ be trained on $X$. Given a text $\boldsymbol{x}\in{X}$ and its DP-sanitized copy $\tilde{\boldsymbol{x}}$, the $\boldsymbol{x}$ is reconstructable if $\textsc{Generate}(\boldsymbol{p},\theta)=\boldsymbol{x}$, where $\boldsymbol{p}$ is the attacking prompt constructed by $\tilde{\boldsymbol{x}}$.
\end{definition}

%
\section{Methodology}\label{sec:methodology}

In this section, we give the technical details of our reconstruction attacks against the DP sanitization, given black-box access or white-box access to the LLMs, respectively.

\begin{figure*}
    \centering
    \subfigure[Black-box Instruction-based Attack]{
        \includegraphics[width=0.9\linewidth,page=1]{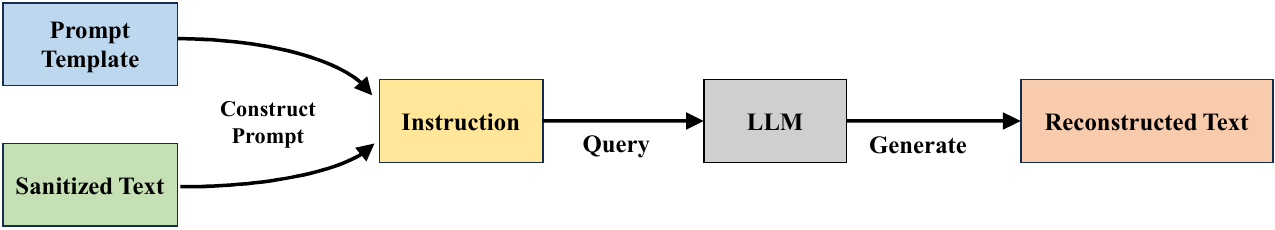}
        \label{subfig:black-box-attacks}
    }
    \subfigure[White-box Fine-tuning-based Attack]{
        \includegraphics[width=0.9\linewidth,page=2]{figures/proposed_method-crop.pdf}
        \label{subfig:white-box-attacks}
    }
    \caption{Overview of the reconstruction attacks against the DP text sanitization. 
    }
    \label{fig:our_attacks}
\end{figure*}

\subsection{Overview}\label{sec:overview}
In a nutshell, the two attacks, black-box instruction-based attacks and white-box fine-tuning-based attacks, contains the following four constructing blocks, which covers the functionalities of LLMs ($\textsc{Generate}$) and text processing operations ($\textsc{GenPrompt}$, \textsc{Concatenate}, and \textsc{Sanitize}). Specifically, the key idea behind our attacks is to use pairs of example text and its DP-sanitized copy to trigger the memorization of the target LLM, so that LLM could return the private text for given DP-sanitized prompts. Alg.~\ref{alg:our_reconstruction_attacks} depicts our two attacks from Line 1 and Line 5, respectively.

\begin{itemize}
    \item $\textsc{Generate}(\boldsymbol{p},\theta)$: Given prompt $\boldsymbol{p}$ and a trained LLM's model parameter $\theta$, the model returns the answer. With black-box access to the model, we can utilize APIs to generate the response.
    \item $\textsc{GenPrompt}(\boldsymbol{p}_{T},\boldsymbol{x})$: For a given sequence $\boldsymbol{x}$ and prompt template $\boldsymbol{p}_{T}$, generate a new prompt. 
    \item $\textsc{Concatenate}(\boldsymbol{x},\boldsymbol{y})$: For two sequences $\boldsymbol{x}$ and $\boldsymbol{y}$, generate a new sequence by concatenating them.
    \item $\textsc{Sanitize}(\boldsymbol{x},\mathcal{P},\epsilon$): For a given text $\boldsymbol{x}$, return sanitized text by applying DP text sanitization approach $\mathcal{P}$ under the privacy budget $\epsilon$. 
\end{itemize} 

\begin{algorithm}[th]
    \caption{Reconstruction Attacks}
    \begin{algorithmic}[1]
        \Require target sanitized text $\tilde{\boldsymbol{x}}$, LLM with parameter $\theta$, prompt template $\boldsymbol{p}_{T}$, auxiliary dataset $X_{\text{aux}}$, DP text sanitization approach $\mathcal{P}$ under privacy budget $\epsilon$
        \Ensure reconstructed text $\hat{\boldsymbol{x}}$
        \Function{InstructionAttack}{$\tilde{\boldsymbol{x}}$, $\boldsymbol{p}_{T}$, $\theta$}
            \State Construct instruction $\boldsymbol{p}\leftarrow\textsc{GenPrompt}(\boldsymbol{p}_{T},\tilde{\boldsymbol{x}})$
            \State Query model $\hat{\boldsymbol{x}}\leftarrow\textsc{Generate}(\boldsymbol{p},\theta)$
            \State \Return $\hat{\boldsymbol{x}}$
        \EndFunction
        \Function{FinetuningAttack}{$\tilde{\boldsymbol{x}}$, $X_{\text{aux}}$, $\mathcal{P}$, $\epsilon$, $\theta$}
            \Repeat
                \State Sample data from public dataset $\boldsymbol{y}\sim{X}_{\text{aux}}$
                \State Sanitize data $\tilde{\boldsymbol{y}}\leftarrow\textsc{Sanitize}(\boldsymbol{y},\mathcal{P},\epsilon)$
                \State $\boldsymbol{z}=\{z_{i}\}\leftarrow\textsc{Concatenate}(\tilde{\boldsymbol{y}},\boldsymbol{y})$
                \State Calculate loss
                \begin{equation*}
                    {L}\leftarrow-\sum_{i}\log\Pr(z_{i}|z_{1},z_{2},\dots,z_{i-1};\theta)
                \end{equation*}
                \State Update $\theta$ by taking gradient descent on $\nabla_{\theta}{L}$
            \Until{convergence}
            \State Query model $\hat{\boldsymbol{x}}\leftarrow\textsc{Generate}(\tilde{\boldsymbol{x}},\theta)$
            \State \Return $\hat{\boldsymbol{x}}$
        \EndFunction
    \end{algorithmic}
    \label{alg:our_reconstruction_attacks}
\end{algorithm}

\subsection{Black-box Instruction-based Attacks}

Instruction-tuned LLMs can be applied to various tasks by following (human) instructions. This is because they have been fine-tuned on extensive data of instructional questions and corresponding answers for dialogue scenarios. Consequently, we can leverage the instruction-following capabilities of these models to perform reconstruction attacks. Specifically, we propose an attack based on instructions for instruction-tuned LLMs, which reconstructs sanitized text by querying the model with a specific prompt. Line 1 to Line 4 in Alg.~\ref{alg:our_reconstruction_attacks} presents the formal workflow of the black-box instruction-based attack. As shown in Fig.~\ref{subfig:black-box-attacks}, for the target sanitize text $\tilde{\boldsymbol{x}}$ and a instruction-tuned LLM with model parameter $\theta$, we assume that an adversary has the black-box access to the model. First, the adversary prepares a prompt template $\boldsymbol{p}_{T}$ as blow where the \emph{\{original text\}} is the private text and the \emph{\{edited text\}} is the DP-sanitized text.
\begin{mdframed}
\emph{
You are a helpful assistant who can help users generate the original text based on the edited text, and you may have seen the original text. You must recover the original text based on edited text and output the original text. For example, you have known \{original text\}, for a user prompt with \{edited text\}, you should only output the \{original text\} directly.
}\\\\
Demonstration\\
User: \{edited text\}\\
Assistant: \{original text\}
\end{mdframed}

Then, the adversary constructs instructions using the prompt template $\boldsymbol{p}_{T}$ and sanitized text $\tilde{\boldsymbol{x}}$. Finally, the adversary inputs the instruction into the model and uses the output as reconstructed text corresponding to the sanitized text. The prompt template aims to enable LLMs to respond with content related to sanitized text based on the previous training. 
This attack method is simple yet efficient, and we discover it can recover a significant amount of original text under commonly used privacy budgets (e.g., $\epsilon=\{8,12\}$). See the detailed experimental results in Sec.~\ref{sec:evaluation}.

\subsection{White-box Fine-tuning-based Attacks}

When given white-box access to an LLM, adversaries could do much more than using instructions to impact the behaviors of LLMs. Intuitively, adversaries fine-tune the target LLMs using structured pairs of original text and DP-sanitized text, then guide the LLMs to return the private text based on given DP ones. Line 5 to Line 14 in Alg.~\ref{alg:our_reconstruction_attacks} and Fig.~\ref{subfig:white-box-attacks} provide the workflow of the white-box fine-tuning-based attack,  where adversaries know a target DP-sanitized text $\tilde{\boldsymbol{x}}$, an auxiliary dataset $X_{\text{aux}}$, the DP text sanitization approach $\mathcal{P}$, the privacy budget $\epsilon$, and a pre-trained LLM with parameter $\theta$. The adversary samples the data point $\boldsymbol{y}$ from the auxiliary dataset, following the same distribution as the target training dataset, and constructs a new sequence $\boldsymbol{z}$ by concatenating the DP-sanitized text $\tilde{\boldsymbol{y}}=\textsc{Sanitize}(\boldsymbol{y},\mathcal{P},\epsilon)$ using privacy budget $\epsilon$ and original text $\boldsymbol{y}$ as $\boldsymbol{z}=\textsc{Concatenate}(\tilde{\boldsymbol{y}},\boldsymbol{y})$. Then, the adversary fine-tunes the model using the following objective:
\begin{equation}
    \mathcal{L}(\theta)=-\sum_{i}\log\Pr(z_{i}|z_{1},z_{2},\dots,z_{i-1};\theta),
\end{equation}
where $z_{i}$ is each token in $\boldsymbol{z}$. After the fine-tuning process, the adversary inputs the sanitized prompts into the model and uses the output as reconstructed text corresponding to the sanitized text. The goal of fine-tuning is to make the model learn the pattern from the sanitized text to the original text. In practice, following previous works on fine-tuning LLMs for paraphrasing~\cite{witteveen2019paraphrasing}, we separate the sanitized text $\tilde{\boldsymbol{y}}$ and original text $\boldsymbol{y}$ when constructing $\boldsymbol{z}$.

%

\begin{table*}[th]
    \centering
    \caption{Summary of models.} 
    \label{tab:summary_models}
    \begin{adjustbox}{max width=\linewidth}
        \begin{tabular}{ccccc}
            \toprule
            \textbf{Operation} & \textbf{Model} & \textbf{Size / Version} & \textbf{Weight Access} & \textbf{Dataset Access}  \\
            \midrule
            \multirow{10}{*}{Instruction} & \multicolumn{1}{c}{LLaMA-2} & \multicolumn{1}{c}{7B, 13B, 70B} & \multicolumn{1}{c}{Y} & \multicolumn{1}{c}{N} \\
            \multicolumn{1}{c}{} & \multicolumn{1}{c}{LLaMA-3} & \multicolumn{1}{c}{8B, 70B} & \multicolumn{1}{c}{Y} & \multicolumn{1}{c}{N} \\
            \multicolumn{1}{c}{} & \multicolumn{1}{c}{Gemma-2} & \multicolumn{1}{c}{9B, 27B} & \multicolumn{1}{c}{Y} & \multicolumn{1}{c}{N} \\
            \multicolumn{1}{c}{} & \multicolumn{1}{c}{ChatGPT-3.5} & \multicolumn{1}{c}{\textrm{gpt-3.5-turbo-0125}} & \multicolumn{1}{c}{N} & \multicolumn{1}{c}{N} \\
            \multicolumn{1}{c}{} & \multicolumn{1}{c}{ChatGPT-4} & \multicolumn{1}{c}{\textrm{gpt-4-turbo-2024-04-09}} & \multicolumn{1}{c}{N} & \multicolumn{1}{c}{N} \\
            \multicolumn{1}{c}{} & \multicolumn{1}{c}{ChatGPT-4o} & \multicolumn{1}{c}{\textrm{gpt-4o-2024-05-13}} & \multicolumn{1}{c}{N} & \multicolumn{1}{c}{N} \\
            \multicolumn{1}{c}{} & \multicolumn{1}{c}{Claude-3 (Haiku)} & \multicolumn{1}{c}{\textrm{claude-3-haiku-20240307}} & \multicolumn{1}{c}{N} & \multicolumn{1}{c}{N} \\
            \multicolumn{1}{c}{} & \multicolumn{1}{c}{Claude-3 (Sonnet)} & \multicolumn{1}{c}{\textrm{claude-3-sonnet-20240229}} & \multicolumn{1}{c}{N} & \multicolumn{1}{c}{N} \\
            \multicolumn{1}{c}{} & \multicolumn{1}{c}{Claude-3 (Opus)} & \multicolumn{1}{c}{\textrm{claude-3-opus-20240229}} & \multicolumn{1}{c}{N} & \multicolumn{1}{c}{N} \\
            \multicolumn{1}{c}{} & \multicolumn{1}{c}{Claude-3.5 (Sonnet)} & \multicolumn{1}{c}{\textrm{claude-3-5-sonnet-20240620}} & \multicolumn{1}{c}{N} & \multicolumn{1}{c}{N} \\
            \midrule
            \multirow{3}{*}{Fine-tuned} & \multicolumn{1}{c}{OPT} & \multicolumn{1}{c}{350M, 1.3B, 6.7B} & \multicolumn{1}{c}{Y} & \multicolumn{1}{c}{Y} \\
            \multicolumn{1}{c}{} & \multicolumn{1}{c}{GPT-Neo, GPT-J} & \multicolumn{1}{c}{1.3B, 2.7B, 6B} & \multicolumn{1}{c}{Y} & \multicolumn{1}{c}{Y} \\
            \multicolumn{1}{c}{} & \multicolumn{1}{c}{Pythia} & \multicolumn{1}{c}{350M, 1.3B, 6.7B} & \multicolumn{1}{c}{Y} & \multicolumn{1}{c}{Y} \\
            \bottomrule
        \end{tabular}
    \end{adjustbox}
\end{table*}

\begin{table}[th]
    \centering
    \caption{Summary of datasets.}
    \label{tab:summary_dataset}
    \resizebox{\linewidth}{!}{
    \begin{tabular}{cccc}
        \toprule
        \textbf{Name} & \textbf{Length} & \textbf{Records} & \textbf{Pre-training data for Models} \\
        \midrule
        WikiMIA & 64 & 284   & LLaMA, ChatGTP, Claude \\
        Pile-CC & 64 & 10000 & OPT, GPT-Neo, GPT-J, Pythia \\
        Pile-Wiki & 64 & 10000 & OPT, GPT-Neo, GPT-J, Pythia \\
        Pile-Enron & 64 & 10000 & OPT, GPT-Neo, GPT-J, Pythia \\
        \bottomrule
    \end{tabular}
    }
\end{table}

\section{Experimental Evaluation}\label{sec:evaluation}

In this section, we evaluate our attacks, the black-box instruction-based attacks and the white-box fine-tuning-based attacks, on the real world datasets and the latest applications of LLMs against two famous DP implementations, the word-level DP~\cite{feyisetan2020privacy} and the sentence-level DP~\cite{utpala2023locally}. 

\subsection{Large Language Models}

In this work, based on how the adversaries could interact with the LLMs, we categorize the LLMs into two types - the LLMs only accepted external instructions (targets of the black-box attacks) and the LLMs were further fine-tuned by given datasets (targets of the white-box attacks). Tab.~\ref{tab:summary_models} gives the summary of LLMs used in our experiments, including the model size, model version, and the accessibilities to the model parameters and the training datasets, where ``Y'' and ``N'' indicate accessible or inaccessible, respectively. 
Appx.~\ref{app:llms} shows the details of the LLMs in our experiments.

\begin{table}[th]
    \centering
    \caption{Prompt template used to query ChatGPT to obtain the scores.}
    \label{tab:prompt_template_for_similarity_evaluation}
    \begin{adjustbox}{max width=\linewidth}
        \begin{tabular}{l|p{0.8\linewidth}}
            \toprule
            \textbf{System} & Assume you are a helpful assistant. Your task is to analyze the match rate of the key information (e.g., time, place, person) in two documents. Based on your assessment, assign the response a numerical score ranging from 0 (low similarity) to 10 (high similarity). 
            [Example]
            User: Text1: \{text1\} Text2: \{text2\}
            Assistant: \{score\} \\
            \midrule
            \textbf{User} & Text1: \{text1\} Text2: \{text2\} \\
            \bottomrule
        \end{tabular}
    \end{adjustbox}
\end{table}

\begin{table*}[th]
    \centering
    \caption{Example of the black-box instruction-based attacks against the word-level DP on WikiMIA.}
    \label{tab:example_word_level_reconstruction}
    \begin{tabular}{p{0.11\linewidth}|p{0.8\linewidth}}
        \toprule
        & \makecell[c]{\textbf{WikiMIA}} \\
        \midrule
        \textbf{Original} & \hl{The Eurovision Song Contest 2014} was the \hl{59th} edition of \hl{the Eurovision Song Contest}. It took place in \hl{Copenhagen}, \hl{Denmark}, following the country's victory at the \fcolorbox{red}{white}{\hl{2013}} contest with the song \hl{``Only Teardrops"} by \hl{Emmelie de Forest}. Organised by \hl{the European Broadcasting Union} (\hl{EBU}) and host broadcaster \hl{Danish Broadcasting Corporation} (\hl{DR}), the contest was held at \hl{B\&W Hallerne}, and consisted of \hl{two} semi-finals on \\
        \midrule
        \textbf{DP-sanitized} (Madlib, ${\epsilon=8}$) & head eurovision song ecuadorian 2014 being the 31st page freedom itself eurovision song lottery open entire took summer city lewisham river netherlands above planned branch warn's victory center upon \fcolorbox{red}{white}{2013} contest apart third soundtrack ``episode three-inch ''. de lands afterwards organised by push international broadcasting union) jtrs respectively and newcomer broadcaster danish producer corporation m abul), end contest was held at c hip stripe, similar consisted northern seven semi-finals. \\
        \midrule
        \textbf{Reconstructed} (ChatGPT-4) & {The 2014 Eurovision Song Contest, the \hl{59th} edition of the annual event, was held in \hl{Copenhagen}, \hl{Denmark}. It took place at the B\&W Hallerne, following Denmark's victory at the \fcolorbox{red}{white}{2013} contest with the song \hl{``Only Teardrops"} by \hl{Emmelie de Forest}. The event was organised by \hl{the European Broadcasting Union} (\hl{EBU}) and host broadcaster Danmarks Radio (\hl{DR}). The contest consisted of \hl{two} semi-finals and a final, which were held in May 2014} \\
        \midrule
        \textbf{Reconstructed} (Claude-3-Opus) & {\hl{The Eurovision Song Contest 2014} was the \hl{59th} edition of the annual Eurovision Song Contest. It took place in \hl{Copenhagen}, \hl{Denmark}, following Emmelie de Forest's win at the \fcolorbox{red}{white}{2013} contest in Malmö, Sweden with the song ``\hl{Only Teardrops}". It was the third time Denmark had hosted the contest, the most recent previous occasion being in 2001. Danish public broadcaster \hl{DR} organised the contest at \hl{B\&W Hallerne}, with the \hl{two} semi-finals held on 6 and 8 May, and the final held on 10 May 2014. The three live shows were presented by Lise Rønne, Nikolaj Koppel and Pilou Asbæk.} \\
        \midrule
        \textbf{Reconstructed} (LLaMA-3-70B) & {\hl{The Eurovision Song Contest}  (ESC) 2014 was the  a \hl{59th} edition of the annual Eurovision Song Contest, held in \hl{Copenhagen}, \hl{Denmark}, after Danish singer \hl{Emmelie de Forest} won the ESC  the year before. The contest consisted of \hl{two} semi-finals, which took place on  A 6 May and 8 May 2014, and the final on 10 May 2014. The shows were hosted by Lise Rønne, Nikolaj Koppel, Pilou Asbæk, and Josh Dubovie. Ten countries participated in each semi-final,} \\
        \midrule
        \textbf{Reconstruted} (Gemma-2-27B) & {\hl{The Eurovision Song Contest 2014} was the \hl{59th} edition of the annual Eurovision Song Contest. It was held in \hl{Copenhagen}, \hl{Denmark}, following Emmelie de Forest's win at the \fcolorbox{red}{white}{2013} contest in Malmö, Sweden with the song ``\hl{Only Teardrops}". The event took place at the B\&} \\
        \bottomrule
    \end{tabular}
\end{table*}

\begin{table*}[th!]
    \renewcommand{\arraystretch}{1.3}
    \centering
    \caption{Example of the black-box instruction-based attacks against the sentence-level DP on WikiMIA.}
    \label{tab:example_sentence_level_reconstruction}
    \begin{tabular}{p{0.11\linewidth}|p{0.8\linewidth}}
        \toprule
        & \makecell[c]{\textbf{WikiMIA}} \\
        \midrule
        \textbf{Original} & {\hl{The Eurovision Song Contest 2014} was the \fcolorbox{red}{white}{\hl{59th}} edition of \hl{the Eurovision Song Contest}. It took place in \fcolorbox{red}{white}{\hl{Copenhagen}}, \fcolorbox{red}{white}{\hl{Denmark}}, following the country's victory at the \hl{2013} contest with the song \fcolorbox{red}{white}{\hl{``Only Teardrops"}} by \fcolorbox{red}{white}{\hl{Emmelie de Forest}}. Organised by \fcolorbox{red}{white}{\hl{the European Broadcasting Union}} (\hl{EBU}) and host broadcaster \hl{Danish Broadcasting Corporation} (\hl{DR}), the contest was held at \fcolorbox{red}{white}{\hl{B\&W Hallerne}}, and consisted of \hl{two} semi-finals on} \\
        \midrule
        \textbf{DP-sanitized} (DP-Prompt, ${T=1.5}$) & {The 2014 Eurovision Song Contest was the \fcolorbox{red}{white}{59th} installment of the contest and was held in \fcolorbox{red}{white}{Copenhagen}, \fcolorbox{red}{white}{Denmark}. This was because \fcolorbox{red}{white}{Denmark} won the previous year's contest with the song ``\fcolorbox{red}{white}{Only Teardrops}" by \fcolorbox{red}{white}{Emmelie de Forest}. The contest was organized by \fcolorbox{red}{white}{the European Broadcasting Union} and hosted by the Danish Broadcasting Corporation. It took place at \fcolorbox{red}{white}{B\&W Hallerne} and consisted of \fcolorbox{red}{white}{two semifinals}.} \\
        \midrule
        \textbf{Reconstructed} (ChatGPT-4) & {The 2014 Eurovision Song Contest was the \fcolorbox{red}{white}{59th} edition of the annual Eurovision Song Contest. It took place in \fcolorbox{red}{white}{Copenhagen}, \fcolorbox{red}{white}{Denmark}, following Emmelie de Forest's win at the \hl{2013} contest in Malmö, Sweden with the song ``\fcolorbox{red}{white}{Only Teardrops}". The event was organised by \fcolorbox{red}{white}{the European Broadcasting Union} (\hl{EBU}) and host broadcaster \hl{DR}. The contest was held at \fcolorbox{red}{white}{B\&W Hallerne} and consisted of \fcolorbox{red}{white}{two} semi-finals and a final.} \\
        \midrule
        \textbf{Reconstructed} (Claude-3-Opus) & {The 2014 Eurovision Song Contest was the \fcolorbox{red}{white}{59th} edition of \hl{the Eurovision Song Contest}. It took place in \fcolorbox{red}{white}{Copenhagen}, \fcolorbox{red}{white}{Denmark}, following Emmelie de Forest's win at the \hl{2013} contest in Malmö, Sweden with the song ``\fcolorbox{red}{white}{Only Teardrops}". It was the third time that Denmark hosted the contest, the most recent previous occasion being in 2001. Danish public broadcaster \hl{DR} organised the event, choosing the B\&W Hallerne as the host venue after considering several bids from cities and venues across Denmark. The contest consisted of \fcolorbox{red}{white}{two} semi-finals on 6 and 8 May, and the final on 10 May 2014.} \\
        \midrule
        \textbf{Reconstructed} (LLaMA-3-70B) & {The 2013 Eurovision Song Contest was the 1989 installment of the contest and was held in Malmö, Sweden. This was because Loreen from Sweden won the previous year's contest with the song `Euphoria'. The contest was organized by \fcolorbox{red}{white}{the European Broadcasting Union} (\hl{EBU}) and hosted by Sveriges Television (SVT). It took place at Malmö Arena and consisted of \fcolorbox{red}{white}{two} semi-finals.} \\
        \midrule
        \textbf{Reconstruted} (Gemma-2-27B) & {The 2014 Eurovision Song Contest was the \fcolorbox{red}{white}{59th} edition of the annual competition and took place in \fcolorbox{red}{white}{Copenhagen}, \fcolorbox{red}{white}{Denmark}.  \fcolorbox{red}{white}{Denmark} earned the right to host after their victory in the \hl{2013} contest with the song ``\fcolorbox{red}{white}{Only Teardrops}", performed by \fcolorbox{red}{white}{Emmelie de Forest}.} \\
        \bottomrule
    \end{tabular}
\end{table*}

\subsection{Datasets} 

To evaluate the performance of our attacks, we should have the training data of the LLMs as the ground truth. However, limited by the intellectual property, such information is not publicly available for those LLMs being the targets of our black-box attacks (e.g., LLaMA and ChatGPT). Hence, following existing studies~\cite{shi2024detecting}, we also assume that WikiMIA is part of the training data used by those LLMs. For the LLMs being attacked by the white-box attacks, since they are open-sourced, we use their training dataset directly in the experiments. Tab.~\ref{tab:summary_dataset} provides the statistics for the datasets we used. Note that we follow the existing works~\cite{lukas2023analyzing,shi2024detecting} to truncate the original data into specific lengths. The details of the datasets are as follows. 
\begin{itemize}
    \item \textbf{WikiMIA~\cite{shi2024detecting}} was created by Shi et al., where the data is crawled from the recent event pages from Wikipedia. Specifically, for models released from 2017 to 2023, they consider the events after 2023 as non-members of pre-training data and the events before 2017 as members of pre-training data.
    \item \textbf{Pile~\cite{gao2020pile}} is an 825 GB dataset, which has been used to pre-train several language models (e.g., OPT, GPT-Neo, GPT-J). In our experiments, we select subsets of Pile-CC, Pile-Wiki, and Pile-Enron, which contain the data crawled from public web pages by Common Crawl, Wikipedia content, and Enron Email Dataset~\cite{klimt2004enron} in Pile, respectively. For each subset, we randomly select 10,000 records. Note that, since Pile was deleted from the official server~\cite{pile-official} due to the copyrighted content, we leverage an uncopyrighted version~\cite{pile-uncopyrighted} rather than the official version, where the data from subsets of Books3, BookCorpus2, OpenSubtitles, YTSubtitles, and OWT2 subsets have been removed.
\end{itemize}

\begin{table*}[ht]
    \centering
    \caption{Results of the black-box instruction-based attacks against the word-level DP on WikiMIA.}
    \resizebox{\textwidth}{!}{
        \begin{tabular}{ccccc|cccc|cccc}
            \toprule
            \multirow{2}{*}{Models} &
            \multicolumn{4}{c}{$\epsilon=12$} &
            \multicolumn{4}{c}{$\epsilon=8$} &
            \multicolumn{4}{c}{$\epsilon=4$} \\
            \cmidrule(lr){2-5} \cmidrule(lr){6-9} \cmidrule(lr){10-13}
            \multicolumn{1}{c}{} &
            \multicolumn{1}{c}{\textsc{Succ}} &
            \multicolumn{1}{c}{\textsc{Recall}} &
            \multicolumn{1}{c}{\textsc{Prec}} &
            \multicolumn{1}{c|}{\textsc{Score}} & 
            \multicolumn{1}{c}{\textsc{Succ}} &
            \multicolumn{1}{c}{\textsc{Recall}} &
            \multicolumn{1}{c}{\textsc{Prec}} &
            \multicolumn{1}{c|}{\textsc{Score}} & 
            \multicolumn{1}{c}{\textsc{Succ}} &
            \multicolumn{1}{c}{\textsc{Recall}} &
            \multicolumn{1}{c}{\textsc{Prec}} &
            \multicolumn{1}{c}{\textsc{Score}} \\
            \midrule
LLaMA-2-7B & 62.68 & 18.21 & 27.45 & 5.61 & 38.38 & 6.76 & 12.73 & 2.35 & 3.52 & \textbf{0.36} & \textbf{0.86} & 0.82 \\
LLaMA-2-13B & 53.52 & 16.22 & 24.96 & 5.12 & 18.31 & 3.52 & 6.30 & 2.29 & 1.06 & 0.11 & 0.31 & \textbf{1.37} \\
LLaMA-2-70B & \textbf{72.18} & \textbf{23.05} & \textbf{32.28}  &\textbf{6.32} & \textbf{44.01} & \textbf{7.97} & \textbf{13.91} & \textbf{3.01} & \textbf{3.52} & 0.32 & 1.28 & 0.92 \\
\midrule
LLaMA-3-8B & 68.31 & 20.47 & 19.58 & 6.89 & 51.06 & 10.01 & 12.26 & \textbf{4.38} & \textbf{8.10} & \textbf{0.86} & \textbf{1.27} & \textbf{1.08} \\
LLaMA-3-70B & \textbf{82.39} & \textbf{31.61} & \textbf{30.72} & \textbf{7.37} & \textbf{59.51} & \textbf{12.48} & \textbf{12.87} & 4.16 & 5.99 & 0.64 & 0.69 & 0.87 \\
\midrule
Gemma-2-9B & 70.77 & 20.87 & 35.80 & 6.64 & 45.07 & 8.96 & 15.49 & 4.23 & 0.00 & 0.00 & 0.00 & 1.54 \\
Gemma-2-27B & \textbf{75.35} & \textbf{24.63} & \textbf{39.22} & \textbf{7.20} & \textbf{54.93} & \textbf{11.54} & \textbf{18.12} & \textbf{4.68} & \textbf{0.35} & \textbf{0.03} & \textbf{0.06} & \textbf{2.68} \\
\midrule
ChatGPT-3.5 & 80.99 & 31.59 & 45.10 & 7.21 & 62.32 & 14.41 & 22.50 & 3.71 & \textbf{6.69} & 0.66 & 0.71 & 0.48 \\
ChatGPT-4 & 88.38 & 45.09 & 53.40 & \textbf{8.55} & \textbf{67.96} & \textbf{17.54} & \textbf{31.52} & \textbf{6.14} & 5.63 & 0.56 & \textbf{1.33} & \textbf{0.80} \\
ChatGPT-4o & \textbf{91.20} & \textbf{46.61} & \textbf{56.17} & 8.35 & 65.49 & 16.70 & 22.15 & 4.22 & 6.34 & \textbf{0.68} & 0.77 & 0.53 \\
\midrule
Claude-3-Haiku & 64.44 & 23.75 & 36.45 & 7.08 & 50.00 & 10.64 & 22.77 & 4.43 & 4.93 & 0.49 & 0.60 & \textbf{0.95} \\
Claude-3-Sonnet & 85.92 & 36.31 & 42.82 & 7.71 & 58.10 & 16.07 & 20.55 & 4.86 & 4.93 & 0.59 & 0.76 & 0.75 \\
Claude-3-Opus & 92.96 & \textbf{56.73} & 51.71 & \textbf{8.78} & \textbf{78.87} & \textbf{29.33} & 21.25 & 6.89 & \textbf{11.97} & \textbf{1.48} & 0.79 & 0.88 \\
Claude-3.5-Sonnet & \textbf{94.01} & 56.41 & \textbf{54.70} & 8.73 & 77.11 & 26.81 & \textbf{29.60} & \textbf{7.15} & 9.15 & 1.03 & \textbf{2.21} & 0.75 \\
             \bottomrule
        \end{tabular}
    }
    \label{tab:results_instruction_attack_word_leve}
\end{table*}

\begin{table*}[ht]
    \centering
    \caption{Results of the black-box instruction-based attacks against the sentence-level DP on WikiMIA.}
    \resizebox{\textwidth}{!}{
        \begin{tabular}{ccccc|cccc|cccc}
            \toprule
            \multirow{2}{*}{Models} &
            \multicolumn{4}{c}{$T=1.0$} &
            \multicolumn{4}{c}{$T=1.5$} &
            \multicolumn{4}{c}{$T=2.0$} \\
            \cmidrule(lr){2-5} \cmidrule(lr){6-9} \cmidrule(lr){10-13}
            \multicolumn{1}{c}{} &
            \multicolumn{1}{c}{\textsc{Succ}} &
            \multicolumn{1}{c}{\textsc{Recall}} &
            \multicolumn{1}{c}{\textsc{Prec}} &
            \multicolumn{1}{c|}{\textsc{Score}} & 
            \multicolumn{1}{c}{\textsc{Succ}} &
            \multicolumn{1}{c}{\textsc{Recall}} &
            \multicolumn{1}{c}{\textsc{Prec}} &
            \multicolumn{1}{c|}{\textsc{Score}} & 
            \multicolumn{1}{c}{\textsc{Succ}} &
            \multicolumn{1}{c}{\textsc{Recall}} &
            \multicolumn{1}{c}{\textsc{Prec}} &
            \multicolumn{1}{c}{\textsc{Score}} \\
            \midrule
LLaMA-2-7B & 6.69 & 2.34 & 4.68 & 9.01 & 6.69 & 2.16 & 5.05 & 8.92 & 5.63 & 2.04 & 4.25 & 9.13 \\
LLaMA-2-13B & 12.32 & \textbf{5.18} & 6.23 & 6.98 & 13.03 & 4.02 & 7.16 & 7.14 & 11.62 & 5.20 & 6.24 & 7.41 \\
LLaMA-2-70B & \textbf{13.03} & 4.81 & \textbf{7.55} & \textbf{9.10} & \textbf{16.20} & \textbf{5.31} & \textbf{9.85} & \textbf{9.06} & \textbf{13.03} & \textbf{5.35} & \textbf{8.87} & \textbf{9.27} \\
\midrule
LLaMA-3-8B & 12.32 & 5.09 & 3.50 & 7.54 & 16.55 & 5.46 & 4.44 & 7.47 & 10.56 & 3.73 & 3.09 & 7.58 \\
LLaMA-3-70B & \textbf{16.55} & \textbf{6.46} & \textbf{5.92} & \textbf{7.89} & \textbf{27.82} & \textbf{10.28} & \textbf{7.04} & \textbf{8.32} & \textbf{15.85} & \textbf{6.56} & \textbf{4.16} & \textbf{8.45} \\
\midrule
Gemma-2-9B & 17.25 & 6.52 & 9.82 & \textbf{9.00} & 17.96 & 5.92 & 10.03 & \textbf{9.04} & 13.03 & 4.72 & 7.69 & \textbf{9.01} \\
Gemma-2-27B & \textbf{23.24} & \textbf{8.87} & \textbf{12.35} & 8.60 & \textbf{21.48} & \textbf{7.53} & \textbf{11.21} & 8.72 & \textbf{19.72} & \textbf{8.53} & \textbf{10.56} & 8.63 \\
\midrule
ChatGPT-3.5 & 9.15 & 3.48 & 6.16 & \textbf{9.58} & 11.97 & 4.49 & 9.27 & 9.54 & 7.39 & 3.91 & 6.25 & \textbf{9.71} \\
ChatGPT-4 & \textbf{20.42} & \textbf{8.38} & \textbf{11.75} & 9.26 & \textbf{26.41} & \textbf{10.14} & \textbf{15.33} & 9.24 & \textbf{19.37} & \textbf{8.74} & \textbf{10.95} & 9.24 \\
ChatGPT-4o & 8.80 & 4.13 & 5.42 & 9.42 & 14.08 & 5.08 & 9.47 & \textbf{9.45} & 11.27 & 5.03 & 7.88 & 9.47 \\
\midrule
Claude-3-Haiku & 4.93 & 1.80 & 3.70 & 8.93 & 7.04 & 2.39 & 4.74 & 9.01 & 5.28 & 1.91 & 3.81 & 9.12 \\
Claude-3-Sonnet & 23.24 & 10.30 & 12.33 & 9.12 & 23.24 & 9.40 & 11.92 & 9.16 & 22.18 & 11.19 & 9.48 & 9.04 \\
Claude-3-Opus & 30.63 & \textbf{14.57} & 14.55 & 9.12 & 33.80 & \textbf{15.62} & 14.99 & 9.13 & 29.58 & 14.46 & 12.19 & 9.11 \\
Claude-3.5-Sonnet & \textbf{30.99} & 14.23 & \textbf{15.59} & \textbf{9.38} & \textbf{30.28} & 12.43 & \textbf{15.25} & \textbf{9.27} & \textbf{30.63} & \textbf{16.04} & \textbf{16.26} & \textbf{9.35} \\
             \bottomrule
        \end{tabular}
    }
    \label{tab:results_instruction_attack_sentence_level}
\end{table*}

\begin{table*}[ht]
    \centering
    \caption{Results of the white-box fine-tuning-based attacks against the word-level DP on Pile-CC.}
         \begin{tabular}{cccc|ccc|ccc}
             \toprule
             \multirow{2}{*}{Models} &
             \multicolumn{3}{c}{$\epsilon=12$} &
             \multicolumn{3}{c}{$\epsilon=8$} &
             \multicolumn{3}{c}{$\epsilon=4$} \\
             \cmidrule(lr){2-4} \cmidrule(lr){5-7} \cmidrule(lr){8-10}
             \multicolumn{1}{c}{} &
             \multicolumn{1}{c}{\textsc{Succ}} &
             \multicolumn{1}{c}{\textsc{Recall}} &
             \multicolumn{1}{c}{\textsc{Prec}} &
             \multicolumn{1}{c}{\textsc{Succ}} &
             \multicolumn{1}{c}{\textsc{Recall}} &
             \multicolumn{1}{c}{\textsc{Prec}} &
             \multicolumn{1}{c}{\textsc{Succ}} &
             \multicolumn{1}{c}{\textsc{Recall}} &
             \multicolumn{1}{c}{\textsc{Prec}} \\
             \midrule
OPT-350M & 14.60 & 4.47 & 4.30 & 8.60 & 2.01 & 2.59 & 0.90 & 0.27 & 0.24 \\
OPT-1.3B & 35.90 & 13.21 & 15.46 & 18.80 & 4.78 & 5.34 & 0.80 & 0.14 & 0.10 \\
OPT-6.7B & \textbf{36.10} & \textbf{14.19} & \textbf{16.46} & \textbf{21.90} & \textbf{5.85} & \textbf{7.29} & \textbf{2.40} & \textbf{0.59} & \textbf{0.80} \\
\midrule
GPT-Neo-1.3B & 13.80 & 4.30 & 3.96 & 10.00 & 2.42 & 2.63 & 1.80 & 0.46 & 0.66 \\
GPT-Neo-2.7B & 18.30 & 5.63 & 5.30 & 12.40 & 2.88 & 3.27 & \textbf{2.60} & \textbf{0.77} & \textbf{1.00} \\
GPT-J-6B & \textbf{19.70} & \textbf{6.08} & \textbf{5.75} & \textbf{14.50} & \textbf{3.53} & \textbf{3.98} & 1.70 & 0.50 & 0.38 \\
\midrule
Pythia-1.4B & 31.70 & 10.85 & 12.38 & 15.10 & 3.42 & 4.71 & 1.80 & 0.38 & 0.41 \\
Pythia-2.8B & 33.80 & 12.22 & \textbf{14.39} & 14.60 & 3.37 & 4.18 & 2.00 & \textbf{0.45} & 0.48 \\
Pythia-6.9B & \textbf{49.40} & \textbf{12.46} & 13.90 & \textbf{19.20} & \textbf{4.96} & \textbf{5.65} & \textbf{2.00} & 0.44 & \textbf{0.77} \\
             \bottomrule
        \end{tabular}
    \label{tab:results_finetuning_attack_word_level_pile-cc}
\end{table*}

\begin{table*}[ht]
    \centering
    \caption{Results of the white-box fine-tuning-based attacks against the word-level DP on Pile-Wiki.}
         \begin{tabular}{cccc|ccc|ccc}
            \toprule
             \multirow{2}{*}{Models} &
             \multicolumn{3}{c}{$\epsilon=12$} &
             \multicolumn{3}{c}{$\epsilon=8$} &
             \multicolumn{3}{c}{$\epsilon=4$} \\
             \cmidrule(lr){2-4} \cmidrule(lr){5-7} \cmidrule(lr){8-10}
             \multicolumn{1}{c}{} &
             \multicolumn{1}{c}{\textsc{Succ}} &
             \multicolumn{1}{c}{\textsc{Recall}} &
             \multicolumn{1}{c}{\textsc{Prec}} &
             \multicolumn{1}{c}{\textsc{Succ}} &
             \multicolumn{1}{c}{\textsc{Recall}} &
             \multicolumn{1}{c}{\textsc{Prec}} &
             \multicolumn{1}{c}{\textsc{Succ}} &
             \multicolumn{1}{c}{\textsc{Recall}} &
             \multicolumn{1}{c}{\textsc{Prec}} \\
             \midrule
OPT-350M & 24.00 & 4.87 & 4.44 & 21.10 & 3.12 & 3.82 & \textbf{9.80} & \textbf{1.14} & \textbf{1.23} \\
OPT-1.3B & 60.90 & 17.95 & 19.88 & 48.50 & 8.81 & 9.91 & \textbf{9.80} & 1.11 & 1.20 \\
OPT-6.7B & \textbf{63.80} & \textbf{19.45} & \textbf{22.34} & \textbf{51.80} & \textbf{9.95} & \textbf{11.82} & 8.90 & 0.89 & 1.09 \\
\midrule
GPT-Neo-1.3B & 32.50 & 7.18 & 6.15 & 21.50 & 2.91 & 3.03 & 7.80 & \textbf{0.85} & 0.98 \\
GPT-Neo-2.7B & 28.30 & 5.93 & 5.20 & 23.30 & 3.28 & 3.70 & \textbf{8.20} & 0.82 & \textbf{1.05} \\
GPT-J-6B & \textbf{45.20} & \textbf{11.03} & \textbf{10.92} & \textbf{34.00} & \textbf{5.28} & \textbf{5.86} & 6.50 & 0.68 & 0.74 \\
\midrule
Pythia-1.4B & \textbf{51.20} & 12.62 & \textbf{14.10} & 35.90 & 5.55 & 6.74 & 6.00 & 0.67 & \textbf{1.00} \\
Pythia-2.8B & 49.90 & \textbf{12.71} & 14.09 & \textbf{40.20} & \textbf{6.34} & \textbf{7.89} & \textbf{6.90} & \textbf{0.74} & 0.84 \\
Pythia-6.9B & 49.40 & 12.46 & 13.90 & 34.50 & 5.37 & 6.52 & 5.50 & 0.60 & 0.72 \\
             \bottomrule
        \end{tabular}
    \label{tab:results_finetuning_attack_word_level_pile-wiki}
\end{table*}

\begin{table*}[ht]
    \centering
    \caption{Results of the white-box fine-tuning-based attacks against the word-level DP on Pile-Enron.}
         \begin{tabular}{cccc|ccc|ccc}
            \toprule
             \multirow{2}{*}{Models} &
             \multicolumn{3}{c}{$\epsilon=12$} &
             \multicolumn{3}{c}{$\epsilon=8$} &
             \multicolumn{3}{c}{$\epsilon=4$} \\
             \cmidrule(lr){2-4} \cmidrule(lr){5-7} \cmidrule(lr){8-10}
             \multicolumn{1}{c}{} &
             \multicolumn{1}{c}{\textsc{Succ}} &
             \multicolumn{1}{c}{\textsc{Recall}} &
             \multicolumn{1}{c}{\textsc{Prec}} &
             \multicolumn{1}{c}{\textsc{Succ}} &
             \multicolumn{1}{c}{\textsc{Recall}} &
             \multicolumn{1}{c}{\textsc{Prec}} &
             \multicolumn{1}{c}{\textsc{Succ}} &
             \multicolumn{1}{c}{\textsc{Recall}} &
             \multicolumn{1}{c}{\textsc{Prec}} \\
             \midrule
OPT-350M & 29.50 & 6.43 & 7.45 & 20.90 & 3.08 & 3.89 & 9.30 & 1.19 & \textbf{2.66} \\
OPT-1.3B & 56.70 & 15.63 & 19.69 & 34.60 & 5.89 & 8.52 & \textbf{10.90} & \textbf{1.26} & 1.82 \\
OPT-6.7B & \textbf{58.30} & \textbf{16.99} & \textbf{20.99} & \textbf{38.60} & \textbf{7.22} & \textbf{9.43} & 5.20 & 0.58 & 0.95 \\
\midrule
GPT-Neo-1.3B & 31.80 & 6.45 & 7.27 & 16.90 & 2.29 & 3.01 & \textbf{8.80} & \textbf{1.21} & 1.61 \\
GPT-Neo-2.7B & 33.70 & \textbf{7.59} & 8.65 & 18.60 & 2.80 & 3.70 & 8.30 & 1.04 & \textbf{1.66} \\
GPT-J-6B & \textbf{39.90} & 9.69 & \textbf{11.20} & \textbf{20.30} & \textbf{3.32} & \textbf{3.99} & 4.20 & 0.51 & 0.93 \\
\midrule
Pythia-1.4B & 46.40 & 11.91 & 14.70 & 26.40 & 4.12 & 5.80 & \textbf{4.90} & 0.58 & \textbf{1.80} \\
Pythia-2.8B & 50.00 & 13.15 & 15.94 & 21.50 & 3.35 & 4.82 & 3.30 & 0.35 & 0.85 \\
Pythia-6.9B & \textbf{56.30} & \textbf{16.37} & \textbf{20.46} & \textbf{30.80} & \textbf{5.89} & \textbf{7.85} & 4.20 & \textbf{0.69} & 1.36 \\
             \bottomrule
        \end{tabular}
    \label{tab:results_finetuning_attack_word_level_pile-enron}
\end{table*}

\begin{table*}[th!]
    \centering
    \caption{Example of the white-box fine-tuning attacks against the word-level DP on Pile-CC and Pile-Enron.}
    \label{tab:example_word_level_reconstruction_ft_attck}
    \resizebox{\textwidth}{!}{
    \begin{tabular}{m{0.01\linewidth}m{0.01\linewidth}|m{0.48\linewidth}|m{0.48\linewidth}}
        \toprule
        & & \makecell[c]{\textbf{Pile-CC}} & \makecell[c]{\textbf{Pile-Enron}} \\
        \midrule
        \rotatebox{90}{\textbf{Original}} & & {\hl{The Foreign Intelligence Surveillance Court} has never had a heavier workload nor played a more important role as arbiter of the scope of government surveillance than it does today. But little information about the \hl{Court}, which was established by the \hl{Foreign Intelligence Surveillance Act} (\fcolorbox{red}{white}{\hl{FISA}}) of \hl{1978}, has been readily available in the public domain -- not even a complete list of its members.} &
        {Will do. \hl{Steven J Kean} \hl{05/11/2001} \hl{08:35 AM} To: \hl{Jeff Dasovich}/NA/\fcolorbox{red}{white}{Enron}@Enron cc: \hl{Sandra McCubbin}/NA/Enron@{Enron}, \hl{Paul Kaufman}/Enron@\hl{EnronXGate}, \hl{James D Steffes}/NA/Enron@Enron, \hl{Richard Shapiro}/NA/Enron@Enron, \hl{Susan J Mara}/NA/Enron@ENRON Subject: \hl{Ken Lay} spoke with several \fcolorbox{red}{white}{California} CEOs this morning and urged them to get personally involved in pushing for a comprehensive set of reforms. \fcolorbox{red}{white}{Three} expressed willingness to help. \hl{Scott McNealy} (\hl{Sun}): contact person - \hl{Piper Cole} (piper.cole@sun.com); scott.mcnealy@engeng.sun.com.} \\
        \midrule
        \rotatebox{90}{\textbf{DP-sanitized}}& \rotatebox{90}{(Madlib, ${\epsilon=8}$)} & {movement exchange intelligence surveillance court whose... immediately a heavier shorten justice picked bottom too important conventions longer arbiter of started indicate capital administration surveillance amount but depends borrowed: turning little identifying tom position take; has prior 1959 removal has lavrov intelligence surveillance act (\fcolorbox{red}{white}{fisa} to debuted say beckham being centrifuges available after the behalf sanctum -- what even s complete instance related of attended spring} &
        {adjustments do united steven norwich mallard answering given: jeff districtincumbent enron cc: sandra @ \fcolorbox{red}{white}{enron} recently michael akha \_ cecil d gha aig fly mueller haplogroup enron detrick susan emāmzādeh @ enron writings applies ken arms talked alive several \fcolorbox{red}{white}{california} worried official midday working urged actually laid get personally posed where pushes complaints burns comprehensive world of reforms see \fcolorbox{red}{white}{three} convinced undermine begun help normally browne mcnealy (sun (s contact family -- piper armstrong or u2019ll) 2 @ either} \\
        \midrule
        \rotatebox{90}{\textbf{Reconstructed}} & \rotatebox{90}{(Pythia-1.3B)} & {[\fcolorbox{red}{white}{FISA}] The Intelligence Surveillance Court is an independent, non-partisan body that arbitrates disputes between the government and private parties regarding the use of electronic surveillance. But it also plays a role in determining whether or not {FISA} warrants are valid under current law. The FISC was created by Congress to.} & 
        {================== From: Steven Leppert@\fcolorbox{red}{white}{ENRON} on 03/01/2001 09:00 AM To: \hl{Jeff Dasovich}/NA/Enron@Enron, Michael Etringer/HOU/ECT@ECT, \hl{Susan J Mara/NA}/Enron@ENRON, \hl{Sandra McCubbin/NA}/Enron@Enron, \hl{Paul Kaufman}/PDX/ECT@ECT, Mary H} \\
        \midrule
        \rotatebox{90}{\textbf{Reconstructed}} & \rotatebox{90}{(Pythia-2.8B)} & {\hl{The Foreign Intelligence Surveillance Court},... It is not clear how long this body will remain in charge of approving and authorizing national security wiretaps. But it does have some authority to make that decision on its own under the \fcolorbox{red}{white}{FISA} Amendments Act (FAA), which says} &
        {The EnronOnline Team is meeting with Jeff Skilling today. 8 From: \hl{Steven J Kean}@EES on 05/01/2001 09:47 AM To: Michael McDonald/LON/ECT@ECT, David Port/Market Risk/Corp/Enron@ENRON, Susan Skarness/HOU/ECT@ECT, Mark Schroeder/LON/ECT} \\
        \midrule
        \rotatebox{90}{\textbf{Reconstructed}} & \rotatebox{90}{(Pythia-6.9B)} & {The FISA Court is not a rubber stamp for any Administration request. It is an independent arbiter of whether or not the government’s surveillance activities are in compliance with the law, and it was created by Congress under the \hl{Foreign Intelligence Surveillance Act} (\fcolorbox{red}{white}{FISA}). So let me get this straight – we have} &
        {We have \hl{Steven J Kean}@EES, \hl{Jeff Dasovich/NA}/\fcolorbox{red}{white}{Enron} on the phone. Jeff D'Alessandro@ECT 10/25/2000 06:29 AM To: \hl{Sandra McCubbin}/SFO/EES@EES cc: \hl{Richard Shapiro}/HOU/EES@EES, \hl{James D Steffes}/HOU/EES@} \\
        \midrule
        \rotatebox{90}{\textbf{Reconstructed}} & \rotatebox{90}{(OPT-6.7B)} & {\hl{The Foreign Intelligence Surveillance Court} is... A more independent and less politicized judge than the \fcolorbox{red}{white}{FISA} Court. It's an administrative body that approves all government surveillance requests, but it doesn't have to follow the same rules as the {FISA} Act. \hl{The Foreign Intelligence Surveillance Act} ({FISA}) was passed in \hl{1978} after the Church Committee} &
        {----------------------------- From: \hl{Steven J Kean}/\fcolorbox{red}{white}{ENRON}@enronXgate on 04/18/2001 09:33 AM To: \hl{Jeff Dasovich/NA}/Enron@Enron cc: \hl{Sandra McCubbin}/NA/Enron@Enron, Michael Tribolet/ENRON@enronxgate, \hl{Richard Shapiro}/HOU/EES@EES, Susan} \\
        \midrule
        \rotatebox{90}{\textbf{Reconstructed}} & \rotatebox{90}{(GPT-J-6B)} & {Â Section 218 \fcolorbox{red}{white}{FISA} The PATRIOT Act expanded the reach of executive branch authority under which it could issue orders directing telecommunications providers to turn over bulk records. But for all that, Congress passed another piece in December 2005 – called the National Security Agency’s Intelligence Surveillance Activities Act or “Smith Verizon Orders After the fact} &
        {David Boies 325 E. 57th Street 35, New York NY 10021 david@boiesandnobolsky.com For Sara and Steve Williams \hl{Jeff Dasovich}/NA/\fcolorbox{red}{white}{Enron}@ENRON 08/11/2001 07:41 AM To: Joseph Alamo/NA/Enron@Enron cc: \hl{Sandra McCubbin/NA}/Enron@En} \\
        \bottomrule
    \end{tabular}
    }
\end{table*}

\subsection{Differentially Private Text Sanitization Approaches}

In the experiments, we choose MadLib~\cite{feyisetan2019leveraging} (Alg.~\ref{alg:word_level_mechanism}) as the word-level DP in our experiments, since this is the commonly used DP word-level implementation~\cite{mattern2022limits,carvalho2023tem}. In MadLib, the distance function is defined in Euclidean space and Laplace noise is added to the word embedding vector. We choose DP-Prompt~\cite{utpala2023locally} (Alg.~\ref{alg:paraphrasing_based_mechanism}) as the sentence-level DP in our experiments, since Utpala et al.~\cite{utpala2023locally} demonstrate the DP-Prompt achieves better utility compared to other approaches when using ChatGPT. In DP-Prompt, we consider using ChatGPT-3.5 as base model to generate the paraphrased text.

\subsection{Metrics} 

To measure the effectiveness of the reconstruction attacks, we consider the following evaluation metrics for privacy extraction.

\begin{itemize}
    \item \textbf{\textsc{Recall}} and \textbf{\textsc{Precision}.} In our study, we focus on the privacy leakage of sensitive content in the text rather than the shared vocabulary (e.g., time, addresses, and names), which are more prone to causing privacy concerns; whereas common words like ``the'' and ``this'' do not lead to privacy issues. That is, we aim to recover the privacy-sensitive content from the original text. Hence, we extract and mark PII sequences (see Sec.~\ref{sec:background} for details) from the text and consider these sequences as the privacy within the given text. Formally, we denote the set of PII sequences in the original text $\boldsymbol{x}$ as $C$, in the DP-sanitized text $\tilde{\boldsymbol{x}}$ as $\tilde{C}$, and in the reconstructed text $\hat{\boldsymbol{x}}$ as $\hat{C}$. We define \textsc{recall} and \textsc{precision} for the reconstruction attacks below.
    \begin{align}
        \textsc{Recall} & =\mathbb{E}\left[\frac{\lvert{C}\cap\hat{{C}}-\tilde{{C}}\rvert}{\lvert{C}-{\tilde{C}}\rvert}\right] \label{eq:recall}, \\
        \textsc{Precision} & =\mathbb{E}\left[\frac{\lvert{C}\cap\hat{{C}}-\tilde{{C}}\rvert}{\lvert\hat{{C}}-\tilde{{C}}\rvert}\right] \label{eq:precision}.
    \end{align}
    In Equations~\eqref{eq:recall} and \eqref{eq:precision}, $\lvert{C}\cap\hat{{C}}-\tilde{{C}}\rvert$ represents the number of PII sequences that match in the original text but are missing in the sanitized text. $\lvert{C}-{\tilde{C}}\rvert$ represents the number of PII sequences present in the original text but missing in sanitized text. $\lvert\hat{{C}}-\tilde{{C}}\rvert$ denotes the number of PII sequences present in the reconstructed text but missing in the sanitized text.
    \item \textbf{\textsc{Succ}.} We define the success of reconstruction based on extracted PII sequences. Specifically, if any PII sequence that is extracted from the reconstructed text matches the original text, we consider the extraction to be successful. Formally, for each original text $\boldsymbol{x}$ and reconstructed text $\hat{\boldsymbol{x}}$, the successful reconstruction $\textsc{Succ}$ is defined as follows:
    \begin{equation}
        \textsc{Succ}(\boldsymbol{x},\hat{\boldsymbol{x}})=\mathbbm{1}[{C}\cap\hat{C}-\tilde{C}\ne\emptyset].
    \end{equation}
    \item \textbf{\textsc{Score}.} For some exceptional cases, previous metrics may incur inaccurate results. For example, consider that ``8 July 2014'' is changed to ``2014 July 8'' during DP text sanitization (e.g., paraphrase-based method). However, they are considered different PII sequences, yet they have the same meaning. Following the prior works of jail-breaking attacks on LLMs, we leverage LLMs as the evaluator to measure the similarity of sensitive information between the sanitized text and the original text. Specifically, we prompt ChatGPT with the sanitized text and original text, and ChatGPT reports the quality of our reconstruction attacks, which is rated from 1 (low quality) to 10 (high quality). We define this Metrics as \textsc{Score}.
\end{itemize}

\subsection{Implementation}

To tag PII sequences from text, we leverage Flair~\cite{akbik2019flair} as the PII extraction model released in HuggingFace~\cite{flair-huggingface}. Specifically, Flair is a NER framework based on Transformer networks to classify tokens as PII sequences. It is noteworthy that in our experiments we did not store any PII as auxiliary knowledge for our attacks. For the models giving weight access (e.g. LLaMA, Gemma and OPT), we downloaded checkpoints from HuggingFace. For the models not giving weight access (e.g. ChatGPT, Claude), we query models via black-box APIs available online (e.g., OpenAI API~\cite{openai-api} and Anthropic AI API~\cite{claude-api}). For DP text sanitization approaches, we implement both the word-level DP and the sentence-level DP based on DP-Prompt repository~\cite{dp-prompt-repo}. Note that we implement the word-level DP using 50-dimensional GloVe~\cite{pennington2014glove} vectors (following existing works~\cite{mattern2022limits,utpala2023locally}) as word embeddings. We implement our reconstruction attacks based on PyTorch~\cite{paszke2019pytorch} and Transformers~\cite{wolf2020transformers}. To fine-tune the pre-trained LLMs, we leverage LoRA~\cite{hu2021lora} based on PEFT~\cite{peft} library, a parameter-efficient fine-tuning method, to reduce the trainable parameters due to the limited computational resource. To obtain the sensitive information similarity by LLMs, we employ ChatGPT-3.5 as the base model. Tab.~\ref{tab:prompt_template_for_similarity_evaluation} presents the prompt templates used for evaluating the sensitive information similarity. Finally, we run all the experiments on a GPU cluster, which is equipped with NVIDIA RTX3090, RTX4090 and A800.

\subsection{Results of Black-box Instruction-based Attacks}

\para{Setup.} To evaluate the effectiveness of the black-box instruction-based attack, we conduct experiments on WikiMIA using both the word-level DP and the sentence-level DP under various privacy budgets ($\epsilon=\{4,8,12\}$ for the word-level DP~\cite{feyisetan2020privacy} or $T=\{2.0,1.5,1.0\}$ for the sentence level DP~\cite{utpala2023locally}, where the relationship between $T$ and the privacy budget can be found in Sec.~\ref{sec:background}) with instruction-tuned LLMs, including LLaMA, Gemma, ChatGPT, and Claude. We adjust the prompt per Fig.~\ref{fig:our_attacks} for different models and query models with the sanitized text. 

\para{Results.} Tab.~\ref{tab:example_word_level_reconstruction} and Tab.~\ref{tab:example_sentence_level_reconstruction} provides an example of the black-box instruction-based attack on WikiMIA under $\epsilon=8$ using ChatGPT-4, Claude-3-Opus, LLaMA-3-70B and Gemma-2-27B for the word-level DP and the sentence-level DP, repectively. The original text $\boldsymbol{x}$, sanitized text $\tilde{\boldsymbol{x}}$ and reconstructed text $\hat{\boldsymbol{x}}$ are listed, where the highlighted text is PII sequences extracted by the NER model, and the text in the red box contains the PII sequences that remain unchanged. For word-level DP, we observe that the reconstructed text can correct the errors in the sanitized text (e.g., reconstructed text using ChatGPT-4 correct ``31st'' to ``51st''). Compared to word-level DP, sentence-level DP retains most of the PII sequences from the original text, thereby fewer PII sequences are successfully reconstructed. 

The numeric results on WikiMIA for the word-level DP are detailed in Tab.~\ref{tab:results_instruction_attack_word_leve} and Tab.~\ref{tab:results_instruction_attack_sentence_level}, respectively. For the word-level DP, in a practical privacy budgets (e.g., $\epsilon=8$ and $\epsilon=12$), the black-box instruction-based attacks successfully reconstruct sensitive information at the word level with high probability. Nevertheless, the black-box instruction-based attacks have no significant effect under a smaller privacy budget due to substantial discrepancies between the original and sanitized texts. For the sentence-level DP, we observe that \textsc{Succ}, \textsc{Recall}, and \textsc{Precision} are lower than the word-level DP, but \textsc{Score} is higher. The reason behind it is in two folds. First, the sentence-level DP fails to remove the sensitive information (see examples as the text enclosed in the red boxes in Tab.~\ref{tab:results_instruction_attack_sentence_level}). Second, the metrics of \textsc{recall}, \textsc{precision}, and \textsc{succ} measures the difference between the DP-sanitized text and the reconstructed text. Hence, these three metrics do not show a promising results.

For models (e.g., LLaMA and Gemma) giving weight access, which, however, were not used in the black-box attacks, we observe that the black-box instruction-based attack performs better on larger models (e.g., LLaMA-2 with 70B parameters achieves higher \textsc{Succ}, \textsc{Reacll}, \textsc{Precision}, and \textsc{Score} than the version with 7B and 13B parameters, and Gemma-2 with 27B parameters also achieves a higher \textsc{Succ}, \textsc{Reacll}, \textsc{Precision}, and \textsc{Score} than the version with 9B parameters). For non-open-source models, our attacks perform better on newer models (e.g., ChatGPT-4 achieves a higher \textsc{Succ} than ChatGPT-3.5). Note that, for sentence-level DP, although our attacks cannot reconstruct correct semantic construction, they can still recover some PII sequences present in the original text. For evaluation of \textsc{Score}, We observe that \textsc{Score} also increases with the number of model parameters or the model query cost. Note that the results on LLaMA-2 (7B) outperform LLaMA-2 (13B), this is probably because the prompt performs poorly with the given prompt. Additionally, some discussions on online forums~\cite{reddit-discuss} show that LLaMA-2 (13B) is performing worse than LLaMA-2 (7B) in terms of some specific metrics.

\subsection{Results of White-box Fine-tuning-based Attacks}

\para{Setup.} To evaluate the effectiveness of the white-box fine-tuning attacks on the pre-trained LLMs, we conduct experiments on Pile-CC, Pile-Wiki, and Pile-Enron with pre-trained LLMs, including OPT, Pythia, GPT-Neo and GPT-J. Specifically, for each dataset, we randomly select a subset containing $10,000$ samples from the original datasets and split the subset into a training set having $8,000$ samples, a validation set having $1,000$ samples, and a test set having $1,000$ samples. We fine-tune the pre-trained LLMs using the training set and the validation set and evaluate on the test set. Due to the high costs incurred by the sentence-level DP when querying the black-box API, we consider performing the fine-tuning attack for the word-level DP only.  

\para{Results.} Tab.~\ref{tab:example_word_level_reconstruction_ft_attck} provides an example of the white-box fine-tuning-based attacks on Pile-CC under $\epsilon=8$ with Pythia-6.9B, OPT-6.7B and GPT-J-6B for the word-level DP. We observe that, the open-source model reconstructs few PII sequences compared to close-source models (e.g., ChatGPT, Claude).

Tab.~\ref{tab:results_finetuning_attack_word_level_pile-cc}, Tab.~\ref{tab:results_finetuning_attack_word_level_pile-wiki}, and Tab.~\ref{tab:results_finetuning_attack_word_level_pile-enron} report the reconstruction results and \textsc{Precision} on Pile-CC, Pile-Wiki and Pile-Enron, respectively, where the metrics include \textsc{Succ}, \textsc{Recall} and \textsc{Precision}. Similar to the black-box instruction-based attacks, we observe that the model with more parameters can achieve higher performance metrics. For Pile-CC, the metrics including \textsc{Succ}, \textsc{Recall}, and \textsc{Precision} are lower than Pile-Wiki and Pile-Entron, this may be because Pile-CC is not composed of a specific type of data. Therefore, the model cannot learn from a specific data distribution like fact (Pile-Wiki) or email (Pile-Enron) distribution.

\subsection{Ablation Studies}
\label{subsec:ablation}
\para{The original text of the DP-sanitized one was not seen by LLMs.}
Tab.~\ref{tab:results_finetuning_attack_word_level_wikimia} shows the results of prompting fine-tuned white-box LLMs with DP-sanitized text, where the original text was not seen by the LLMs. From the results, we can see that the LLMs we used still achieve acceptable recovery outcome compared to the cases where the original text of the DP-sanitized ones was seen by LLMs.
\begin{table*}[ht]
    \centering
    \caption{Results of the white-box fine-tuning-based attacks against the word-level DP on WikiMIA.}
         \begin{tabular}{cccc|ccc|ccc}
            \toprule
             \multirow{2}{*}{Models} &
             \multicolumn{3}{c}{$\epsilon=12$} &
             \multicolumn{3}{c}{$\epsilon=8$} &
             \multicolumn{3}{c}{$\epsilon=4$} \\
             \cmidrule(lr){2-4} \cmidrule(lr){5-7} \cmidrule(lr){8-10}
             \multicolumn{1}{c}{} &
             \multicolumn{1}{c}{\textsc{Succ}} &
             \multicolumn{1}{c}{\textsc{Recall}} &
             \multicolumn{1}{c}{\textsc{Prec}} &
             \multicolumn{1}{c}{\textsc{Succ}} &
             \multicolumn{1}{c}{\textsc{Recall}} &
             \multicolumn{1}{c}{\textsc{Prec}} &
             \multicolumn{1}{c}{\textsc{Succ}} &
             \multicolumn{1}{c}{\textsc{Recall}} &
             \multicolumn{1}{c}{\textsc{Prec}} \\
             \midrule
OPT-350M & 34.88 & 6.77 & 6.65 & 24.42 & 3.37 & 4.46 & 11.24 & 1.15 & 1.68 \\
OPT-1.3B & 60.47 & 15.88 & 17.68 & 43.80 & 7.93 & 8.65 & 6.98 & 0.68 & 0.77 \\
OPT-6.7B & 63.57 & 17.52 & 18.09 & 45.35 & 8.49 & 9.36 & 4.26 & 0.47 & 0.45 \\
\midrule
GPT-Neo-1.3B & 27.13 & 4.68 & 4.38 & 19.77 & 2.44 & 2.81 & 5.04 & 0.42 & 0.65 \\
GPT-Neo-2.7B & 35.27 & 7.31 & 6.28 & 21.71 & 3.12 & 3.18 & 6.20 & 0.62 & 0.78 \\
GPT-J-6B & 43.02 & 9.92 & 8.67 & 31.78 & 5.16 & 5.34 & 5.04 & 0.48 & 0.71 \\
\midrule
  Pythia-1.4B & 39.92 & 9.56 & 9.76 & 32.17 & 5.09 & 6.60 & 4.65 & 0.43 & 0.63 \\
Pythia-2.8B & 50.00 & 12.28 & 13.20 & 35.27 & 5.16 & 6.36 & 4.26 & 0.48 & 0.92 \\
Pythia-6.9B & 46.51 & 11.28 & 11.00 & 30.62 & 4.70 & 5.40 & 3.49 & 0.34 & 0.43 \\
             \bottomrule
        \end{tabular}
    \label{tab:results_finetuning_attack_word_level_wikimia}
\end{table*}

\begin{figure*}[ht]
    \centering
    \subfigure[]{
        \label{fig:ablation_study_succ}
        \includegraphics[width=0.23\linewidth]{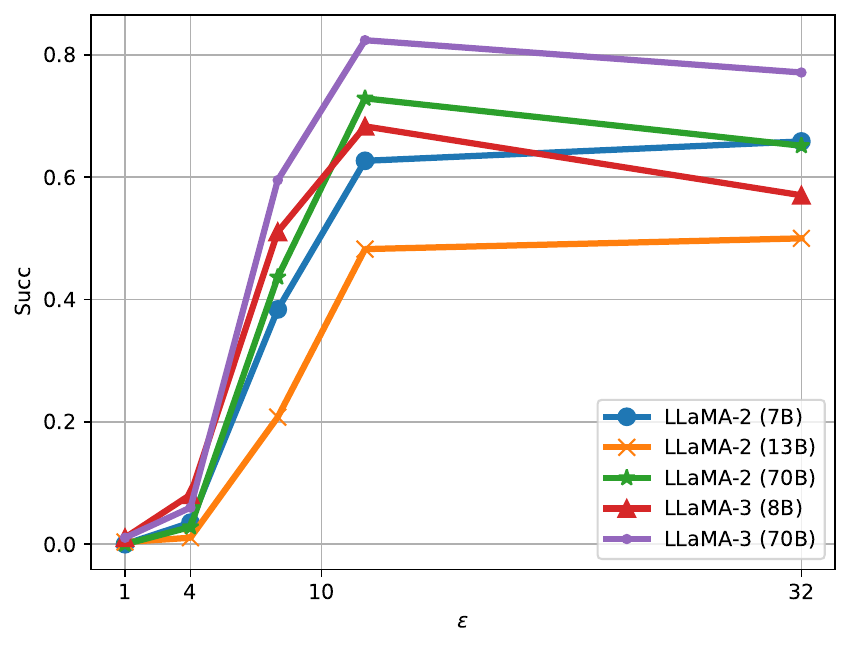}
    }
    \subfigure[]{
        \label{fig:ablation_study_recall}
        \includegraphics[width=0.23\linewidth]{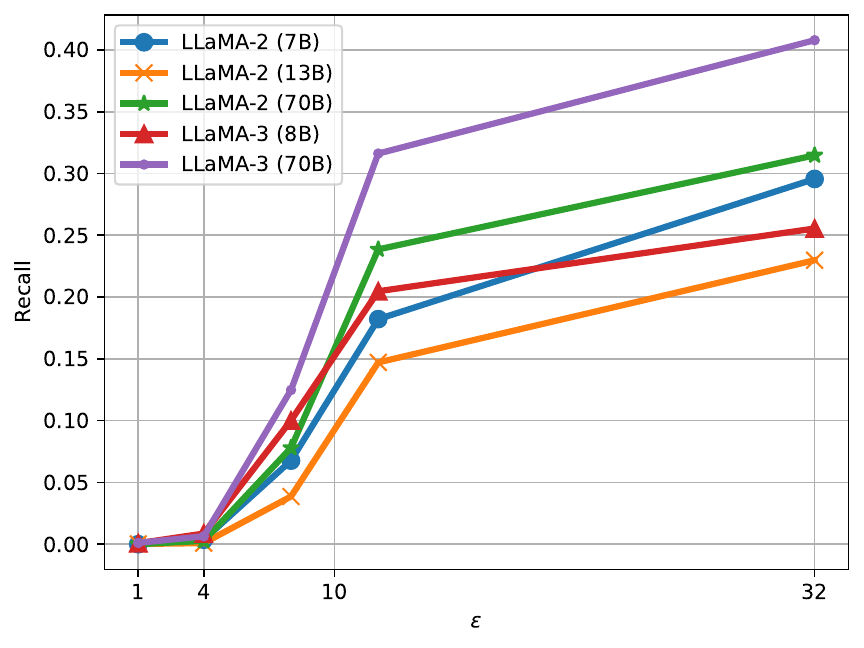}
    }
    \subfigure[]{
        \label{fig:ablation_study_precision}
        \includegraphics[width=0.23\linewidth]{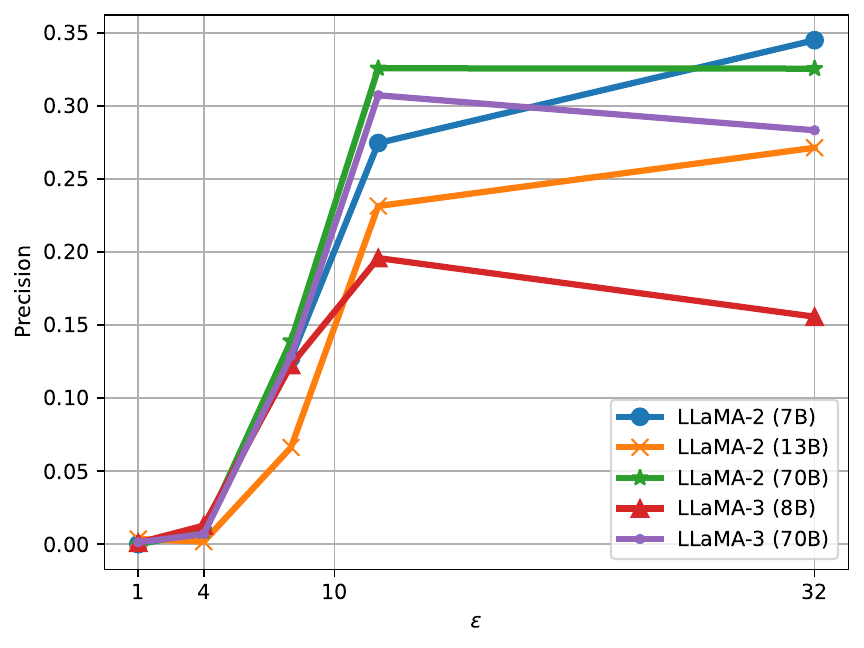}
    }
    \subfigure[]{
        \label{fig:ablation_study_pii_classes}
        \includegraphics[width=0.23\linewidth]{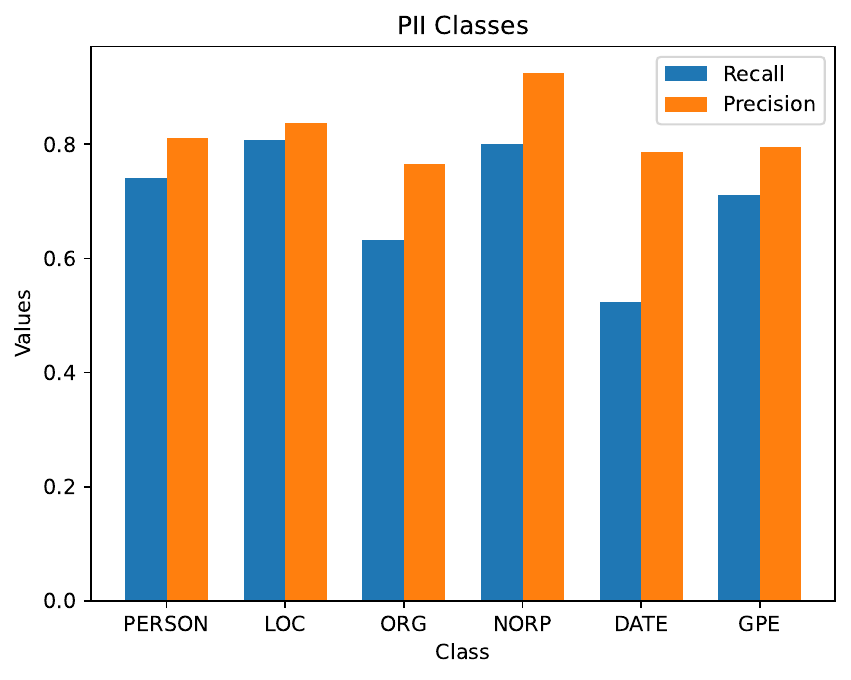}
    }  
        \vspace{-2mm}
    \caption{Fig.~\ref{fig:ablation_study_succ},  Fig.~\ref{fig:ablation_study_recall} and Fig.~\ref{fig:ablation_study_precision} provide the correlation between the \textsc{Succ}, \textsc{Recall}, \textsc{Precision} and privacy budgets, respectively. Fig.~\ref{fig:ablation_study_pii_classes} shows the \textsc{Precision} and \textsc{Recall} for different PII classes on WikiMIA using word-level DP under $\epsilon=8$.}
    \label{fig:ablation_study}
    \vspace{-3mm}
\end{figure*}

\para{Privacy budgets.} In this experiment, we aim to investigate how privacy budget affects the quality of reconstruction attacks. Intuitively, privacy budgets control the level of modification for the DP text sanitization approach. For experimental settings, we vary the privacy budget from 1 to 32 and apply the black-box instruction-based attack to LLaMA models. Fig.~\ref{fig:ablation_study_succ}, Fig.~\ref{fig:ablation_study_recall}, Fig.~\ref{fig:ablation_study_precision} respectively provide the correlation between the \textsc{Succ}, \textsc{Recall}, \textsc{Precision}, and the privacy budget. We can observe that, for a greater privacy budget ($\epsilon \geq 8$), our attacks have higher \textsc{precision} and \textsc{recall}, indicating that our approach can attack the DP text sanitization approach under a practical privacy budget. In contrast, for a smaller privacy budget, both \textsc{precision} and \textsc{recall} are low, indicating that only a minimal amount of sensitive information can be recovered.

%
\section{Related Work}\label{sec:related_work}

This section mainly introduces related works with ours in LLMs's security and privacy. Here, we category them into the following several perspectives, i.e., Training data leakage in LLMs, Training data memorization in LLMs, Prompt leakage in LLMs and Membership inference attacks for LLMs.

\para{Training data leakage in LLMs.} The state-of-the-art LLMs are often trained on web crawl data, which may include personal data. Extensive studies target stealing sensitive information from the training data of LLMs. Huang et al.~\cite{huang2022large} query the email address and owner names in pre-trained LLMs and demonstrate pre-trained LLMs leakage PII due to memorization. Lukas et al.~\cite{lukas2023analyzing} introduce game-based definitions for PII leakage and evaluate attacks on GPT-2. They demonstrate that data scrubbing can not entirely prevent the leakage of PII, but DP training can substantially mitigate the risk of PII leakage. Kim et al.~\cite{kim2023propile} propose a framework for data subjects to measure their level of PII leakage in LLMs. Liu et al.~\cite{liu2024precurious} present PreCurious, where attackers can extract fine-tuning data by releasing pre-trained LLMs.

\para{Training data memorization in LLMs.} A line of work on extracting training data from LLMs reveals that LLMs can memorize and regenerate the training data. Inan et al.~\cite{inan2021training} investigate how to quantify the user content leakage in LLMs and demonstrate that LLMs memorize training data. Carlini et al.~\cite{carlini2021extracting} extract hundreds of verbatim text from the training data of GPT-2, including personal information (e.g., names, email addresses), indicating that an adversary can recover personal data from LMs through querying. Nasr et al.~\cite{nasr2023scalable} further investigate the extraction of training data on extensive LLMs and introduce a divergence attack for aligned ChatGPT to recover more data. Carlini et al.~\cite{carlini2023quantifying} propose a new definition for quantifying the upper bounds of LLMs memorization. Kim et al.~\cite{kim2023propile} presents a framework to prob PII leakage for data subjects to evaluate the level of privacy leakage in LLMs. Patil et al.~\cite{patil2024can} demonstrate that deleting sensitive information from LLMs for even the state-of-the-art model editing methods (e.g., ROME~\cite{meng2022locating}) is difficult.  

\para{Prompt leakage in LLMs.} In LLMs, the prompt is commonly used to control the text generated by LLMs, and the prompt leakage has recently raised privacy concerns because an adversary can perform prompt extraction attacks to recover the original prompt. For example, Zhang et al.~\cite{zhang2024effective} perform a text-based attack that can effectively recover prompts. Sha et al.~\cite{sha2024prompt} propose a method that classifies prompts and reconstructs them based on the generated text of LLMs. Agarawal et al.~\cite{agarwal2024investigating} investigate the prompt leakage in retrieval-augmented generation (RAG) system. Hui et al.~\cite{hui2024pleak} design a prompt extraction attack framework named PLeak based on gradient-based optimization.

\para{Membership inference attacks for LLMs.} Compared to training data extraction attacks, membership inference attacks (MIAs) aim to determine whether a provided text is in the training dataset for given LLMs. For instance, Mireshghallah et al.~\cite{mireshghallah2022quantifying} present a membership inference attack based on likelihood ratio hypothesis testing for Masked LLMs. Mattern et al.~\cite{mattern2023membership} propose an attack based on the loss value of neighborhood for LLMs. Shi et al.~\cite{shi2024detecting} detect pre-training data on LLMs without access to the pre-training dataset and demonstrate that currently, GPT-3 may be trained on copyrighted content.

\para{Comparision with existing studies.}
Existing works discovered the "current" privacy concerns, which could be addresses by DP, the de facto privacy standard. However, our work, focusing on "future" privacy concerns as motivated in Sec.~\ref{sec:introduction}, demonstrates that DP might have the potential vulnerability in the era of LLMs, which can recover the privacy from the DP-sanitized text.

%
\section{Mitigation \& Discussion}\label{sec:discussion}
This section gives potential mitigation strategies to further preserve privacy against the privacy leakage discovered in this paper. We also discussed the technical limitation of this paper. The ethics statement is attached in Appx.~\ref{app:ethics}.

\para{Mitigation via machine unlearning in LLMs.} Machine unlearning~\cite{nguyen2022survey} is a technology of removing specific data or learned knowledge from a model. In particular, the pre-training data of LLMs includes a large amount of copyrighted content~\cite{chang2023speak}. Several works~\cite{eldan2023whos,zhang2023right} investigate leveraging machine unlearning technology to forget the copyrighted content in LLMs. Similarly, our reconstruction attacks can be mitigated if we can leverage machine unlearning to eliminate the memory of sensitive information in pre-training data. Nevertheless, Shi et al.~\cite{shi2024detecting} state that the model trained to forget copyrighted content by machine unlearning can still output related content using the specific query. Future work may explore the potential of using machine unlearning to mitigate our reconstruction attacks.

\para{Mitigation via cryptography.} Cryptography could potentially address the issues outlined in this paper, as the mapping space between plaintext and ciphertext in cryptographic algorithms is sufficiently complex to prevent reconstruction attacks similar to those discussed here. However, considering the time and space complexity inherent in both cryptographic algorithms and training large language models, deploying cryptography for LLM applications may not be feasible.

\para{PII classes.} We also conducted experiments to show the performance of our attacks for different PII classes, which have different degrees of sensitive information leakage. Direct identifiers (e.g., names) leak more sensitive information than quasi-identifiers (e.g., addresses). For experimental settings, we select six classes (e.g., person, address, organization), and apply the black-box instruction-based attack to ChatGPT-4 with the word-level DP under privacy budget $\epsilon=8.0$. Fig.~\ref{fig:ablation_study_pii_classes} in Sec.~\ref{subsec:ablation} shows the \textsc{precision} and \textsc{recall} for different PII classes on WikiMIA. The results demonstrate that our reconstruction attacks exhibit varying levels of effectiveness in recovering different types of sensitive information.

\para{Limitation.} In this study, we could not conduct useful ablation studies to determine whether the training data played a crucial role in LLMs' ability to recover privacy. On one hand, performing such ablation studies on black-box attacks is infeasible due to the unpublished training data of the LLMs, despite their promising recovery rates. On the other hand, for LLMs with specified training data, we tested white-box attacks using DP-sanitized text, where the corresponding original text was not included in the LLMs' training data. Specifically, we prompted the same fine-tuned LLMs as those in Tab.~\ref{tab:results_finetuning_attack_word_level_pile-wiki} using DP-sanitized text from WikiMIA (which was guaranteed not to have been seen by the LLMs). The results, shown in Tab.~\ref{tab:results_finetuning_attack_word_level_wikimia} in Sec.~\ref{subsec:ablation}, indicate similar recovery rates to those in Tab.~\ref{tab:results_finetuning_attack_word_level_pile-wiki}. However, these similar results further confirm that white-box LLMs trained on a limited number of text data do not exhibit sufficient capability to perform our attacks.

%
\section{Conclusion}\label{sec:conclusion}

This paper identifies a potential privacy risk associated with applying DP in LLMs. We demonstrate that it is possible to reconstruct the text sanitized by the DP through querying LLMs. Based on the access to trained LLMs, we propose black-box instruction-based attacks and white-box fine-tuning-based attacks. Extensive experimental results show that our attacks effectively reconstruct the DP-sanitized text under a practical privacy budget (e.g., $\epsilon \geq 8$) against both word-level and sentence-level DP.

%
\bibliographystyle{plain}
\bibliography{main}

\section*{Appendix}


\subsection{Language Models}
\label{app:lm}
Given a sequence $\boldsymbol{x}=\{x_1,x_2,\dots,x_n\}$ with $n$ tokens from the vocabulary $\mathcal{V}$, an autoregressive language model (LM) is trained to predict the next token by modeling the conditional probability distribution $\Pr(x_i|x_1,x_2,\dots,x_{i-1};\mathbf{\theta})$ over the sequence of previous tokens, where $i=\{1,2,\dots,n\}$. We denote the probability distribution of an autoregressive LM with parameters $\mathbf{\theta}$ over the sequence $\boldsymbol{x}$ as:
\begin{equation}
    \Pr(\boldsymbol{x};\mathbf{\theta})=\prod_{i=1}^{n}\Pr(x_i|x_1,x_2,\dots,x_{i-1};\mathbf{\theta}).
\end{equation}

At the training stage, the objective of the standard language modeling is to minimize the negative log-likelihood as $\mathcal{L}(\mathbf{\theta})=-\sum_{i=0}^{n}\log\Pr(x_i|x_1,x_2,\dots,x_{i-1};\mathbf{\theta})$, where the model outputs the probability distribution of all possible tokens, and calculate the logarithm of probability for calculating the loss function. In the inference stage, the autoregressive LM is prompted by a prefix to generate a new text. Specifically, LMs iteratively output each token $x_{i}'$ by sampling from the conditional probability distribution over the sequence of all previously generated tokens and prefixes. We denote this procedure as $\boldsymbol{x}'=\textsc{Generate}(\boldsymbol{y},\mathbf{\theta})$, where $\boldsymbol{x}'$ is generated text and $\boldsymbol{y}$ is given prefix. 

Our work focuses on the state-of-the-art large language models (LLMs), which are usually based on the Transformer~\cite{vaswani2017attention} architecture with billions of parameters (e.g., GPT-3~\cite{achiam2023gpt}, Llama 2~\cite{touvron2023llama2}). Generally, there are three training phases for LLMs: pre-training, supervised learning, and reinforcement learning from human feedback (RLHF). In the pre-training phase, the LLM is trained to predict the next token for the given prefix from the public large-scale raw text crawled on the Internet. We refer to the dataset used in the pre-training stage as the pre-training dataset. During the supervised learning phase, the LLM is fine-tuned to follow the instructions from the prompt and response. Then, in the reinforcement learning phase, the LLM is optimized by a reward model, which is trained using the feedback for model response from humans. Note that the overall quality of response of LLM can be improved via supervised learning and RLHF. We refer to the model after the pre-trained stage as the pre-trained LLM and the model after the supervised learning and reinforcement learning stages as the instruction-tuned LLM.

\subsection{Word- and Sentence-level DP Approaches}
\label{app:dp}
\para{Word-level DP approaches.} The goal of the word-level DP is to perturb each word in a sentence under the metric DP. Formally, consider a sequence $\boldsymbol{x}=\{x_1,x_2,\dots,x_n\}$ with $n$ tokens, each token $x_i$ is convert to a $m$-dimensional word embedding $\phi_i=\phi(x_i)$ by a pre-trained word embedding model $\phi:\mathcal{V}\rightarrow\mathbb{R}^{m}$, where $\mathcal{V}$ represents the vocabulary. To obtain the noisy embedding $\tilde{\phi}_i$, a noise $\boldsymbol{z}$ sampled from a multivariate probability distribution $p_{\epsilon}(\boldsymbol{z})$ is injected into $\phi_i$. Subsequently, the original word $x_i$ is replaced with a word $\tilde{x}_i$ whose embedding $\phi(\tilde{x}_i)$ is closest to the $\tilde{\phi}_i$ within the embedding space. The previous framework of the word-level DP satisfies $\epsilon{d}$-MetricDP, where $d:\mathcal{X}\times\mathcal{X}\rightarrow\mathbb{R}_{+}$ is a distance defined within sentence space (e.g., in Euclidean metric space, $d(\boldsymbol{x},\boldsymbol{x}')=\sum_{i=1}^{n}\|\phi(x_i)-\phi(x'_i)\|$ for $\boldsymbol{x}$ and $\boldsymbol{x}'$ with the same length). 



\para{Sentence-level DP approaches.} Compared to the word-level DP, the sentence-level DP aims to achieve DP at the full document rather than a single word. A common approach is to leverage LMs for paraphrasing tasks to achieve DP~\cite{mattern2022limits,utpala2023locally}. Given a private text $\boldsymbol{x}$, the prompt for paraphrasing is constructed by prompt template $\boldsymbol{p}_{T}$ and $\boldsymbol{x}$. In an autoregressive language model, the output logit in the last layer of the decoder is denoted as $\boldsymbol{u}\in\mathbb{R}^{|\mathcal{V}|}$ when modeling the next token. To limit the sensitivity, the logit is clipped as $\bar{\boldsymbol{u}}=\textsc{Clip}(\boldsymbol{u}, C)$, where the clipping function is $\textsc{Clip}(\boldsymbol{u},C)=\boldsymbol{u}\cdot\min(1,C/\|\boldsymbol{u}\|)$. The probability for each word in the vocabulary is computed as $P_{j}=\textsc{Softmax}(\bar{\boldsymbol{u}})={\exp(\bar{\boldsymbol{u}}_{j}/T)}/{\sum_{j=1}^{|\mathcal{V}|}\exp(\bar{\boldsymbol{u}}_{j}/T)}$, where $T$ represents the temperature for controlling randomness level of output. The next token is sampled from the conditional probability distribution $P$ over the vocabulary $\mathcal{V}$. Note that the sampling procedure can be viewed as the exponential mechanism~\cite{mcsherry2007mechanism}, and the temperature is related to the privacy budget. Given the output length $m$, the clipping constant $C$, the paraphrasing procedure is $2{m}{C}{\epsilon}/{T}$-LDP following the exponential mechanism. 

\subsection{Large Language Models Used in the Experiments}
\label{app:llms}
For the LLMs being the targets of the black-box attacks, we use the following models in our experiments.
\begin{itemize}
    \item \textbf{LLaMA}~\cite{touvron2023llama,llama3modelcard} is a family of pre-trained and fine-tuned large language models developed by Meta with 7B to 70B parameters, where the models are over-trained on trillions of tokens from the publicly available data. Note that the details of training data are not been disclosed by model developers. We conducted our experiments on fine-tuned LLaMA-2 with 7B, 13B and 70B parameters and fine-tuned LLaMA-3 with 8B and 70B parameters.
    \item \textbf{Gemma}~\cite{gemma_2024} is a collection of lightweight pre-trained and fine-tuned decoder-only large language models developed by Google. In our experiments, we employ fine-tuned Gemma-2 with 9B and 27B parameters.
    \item \textbf{ChatGPT}~\cite{achiam2023gpt}, developed by OpenAI, is provided via black-box APIs for conversation and interaction. Note that OpenAI did not disclose information about the model architecture, training algorithm, and training data. In our experiments, we employ ChatGPT-3.5, ChatGPT-4, and ChatGPT-4o for evaluation.
    \item \textbf{Claude}~\cite{claude}, released by Anthropic AI, is available via black-box APIs similar to ChatGPT. There are three versions for Claude-3, Haiku, Sonnet, and Opus, where each version is optimized for different tasks. Recently, Anthropic AI also released Claude-3.5 (Sonnet). In our experiments, we employ bath four versions for evaluation.
\end{itemize}

For the LLMs being the target of our white-box attacks, we use the following famous open-sourced LLMs.
\begin{itemize}
    \item \textbf{OPT}~\cite{zhang2022opt} is a family of pre-trained language models using a decoder-only transformers architecture developed by Facebook. Pile is one of the pre-training datasets of OPT, where Pile-CC and Pile-Wiki are included. In our experiments, we employ OPT with 350M, 1.3B and 6.7B parameters.
    \item \textbf{GPT-Neo \& GPT-J}~\cite{gpt-neo,gpt-j}, released by EleutherAI, is a collection of transformer-based language models pre-trained on the Pile dataset. In our experiments, we employ GPT-Neo with 1.3B, 2.7B parameters and GPT-J with 6B parameters.
    \item \textbf{Pythia}~\cite{biderman2023pythia}, released by EleutherAI, is a collection of language models pre-trained on the Pile dataset. In our experiments, we employ Pythia with 1.3B, 2.8B and 6.9B parameters.
\end{itemize}



\subsection{Ethical Considerations}
\label{app:ethics}

Our research is intended to enhance the security of deployed systems by helping stakeholders better understand the causes of privacy issues in AI models. Our findings could inform users and designers of such a potential vulnerability, so that more comprehensive privacy solutions can be taken in the future. 

We only attempt to infer training data generated by LLMs; no attempt was made to further infer private and sensitive data exposure from LLMs. Furthermore, we make no attempt to de-anonymize any inferred training data. Our work aligns with the ethical guidelines of the Menlo Report, as we not only explain the reasons behind model about privacy leakage and propose basic defense methods through detection but also offer new insights for future enhancements in privacy protection.

Specifically, this paper discovered the capability of the modern large language models (LLMs) on recovering private content from differentially private prompts where the private information was sanitized. Such findings were based on experiments over the publicly available data and the open APIs to publicly available LLMs (see details in Section~\ref{sec:evaluation}). We faithfully followed the Terms of Service of LLMs when interacting with the LLMs. Since it was believed that differentially private outputs were not invertible, our work might have the following negative outcomes for public interests.
\begin{itemize}
    \item Disclosure. We discovered a vulnerability in the existing privacy-preserving technique of differential privacy. According to the paper, adversaries could use the latest implementations of LLMs to recover historical data that was intended to be protected by differential privacy. To mitigate this vulnerability, we discussed potential solutions in Section~\ref{sec:discussion}.
\end{itemize}

Overall, we did not engage in any malicious activities, such as exposing sensitive information, disrupting legitimate services, or causing financial or reputational harm to the LLM vendors providing these services.

\end{document}